\documentclass[twocolumn,showpacs,showkeys,preprintnumbers,amssymb,aps,superscriptaddress,pre]{revtex4-2}
\usepackage{graphicx}
\usepackage{dcolumn}
\usepackage{bm}
\usepackage{epsfig}
\usepackage{color}

\begin{document}

\title{Towards lattice-gas description of low-temperature properties above the Haldane and cluster-based Haldane ground states of a mixed spin-(1,1/2) Heisenberg octahedral chain
}
\author{Katar\'ina Karl'ov\'a} 
\email{katarina.karlova@upjs.sk}
\affiliation{Institute of Physics, Faculty of Science, P. J. \v{S}af\'{a}rik University, Park Angelinum 9, 04001 Ko\v{s}ice, Slovakia}
\author{Jozef Stre\v{c}ka}
\affiliation{Institute of Physics, Faculty of Science, P. J. \v{S}af\'{a}rik University, Park Angelinum 9, 04001 Ko\v{s}ice, Slovakia}
\author{Johannes Richter} 
\affiliation{Institut f\"ur Physik, Otto-von-Guericke Universit\"at in Magdeburg, 39016 Magdeburg, Germany}
\affiliation{Max-Planck-Institut f\"{u}r Physik Komplexer Systeme,
        N\"{o}thnitzer Stra{\ss}e 38, D-01187 Dresden, Germany}

\date{\today}

\begin{abstract}
The rich ground-state phase diagram of the mixed spin-(1,1/2) Heisenberg octahedral chain was previously elaborated from  effective mixed-spin Heisenberg chains, which were derived by employing a local conservation 
of a total spin on square plaquettes of an octahedral chain. 
Here we present a comprehensive analysis of the thermodynamic properties of this
model. In the highly-frustrated parameter region the lowest-energy 
eigenstates of the mixed-spin Heisenberg octahedral chain belong to flat bands, which allow a precise description 
of low-temperature magnetic properties within the localized-magnon approach exploiting a 
classical lattice-gas model of hard-core monomers. The present article provides a more comprehensive version 
of the localized-magnon approach, which extends the range of its validity down to a less frustrated parameter region involving the Haldane and  cluster-based Haldane ground states. 
A comparison 
between results of the developed localized-magnon theory and
accurate numerical methods such as full exact diagonalization and finite-temperature Lanczos technique convincingly 
evidence that the low-temperature magnetic properties above the Haldane and 
the cluster-based Haldane ground states can be extracted from a classical lattice-gas model of hard-core monomers and dimers, which is additionally supplemented by a hard-core particle spanned over the whole lattice representing the gapped Haldane phase.
\end{abstract}


\maketitle

\section{Introduction}
Electron spin systems represent promising candidate for a design of quantum computers, because a two-level character of the electron spin provides one of the simplest platforms to encode a quantum bit \cite{15aaa}. However, the loss of quantum information due to a quantum decoherence is regarded as the most principal obstacle for the development of all quantum technologies exploiting solid-state materials \cite{16aaa}.  Molecular magnetic materials, which are composed from discrete magnetic molecules, belong to the most perspective electron spin systems for quantum computation and quantum information processing \cite{18aaa,19aaa}. A targeted design of molecular magnets through a chemistry-based bottom-up approach ensures their scalability, which allows not 
only the implementation of a single qubit but also greater number of qubits integrated into a more complex quantum circuit that can store and process quantum information \cite{20aaa}. The molecular magnetic materials generally possess
a well defined pattern of discrete energy levels, whereby the associated quantum states can be easily tuned and coherently manipulated \cite{17aaa}. A coherence time of the molecular magnets has been also significantly enhanced by suppressing a quantum decoherence arising mostly from nuclear spins and dipolar forces \cite{21aaa}. The molecular magnetic materials thus naturally satisfy most important requirements imposed on basic building blocks of quantum computers \cite{12aaa,13aaa} when proving their usefulness as prominent resources for the quantum computation \cite{22aaa}, the storage and processing of quantum information \cite{23aaa}. 

The concept of localized magnons \cite{1schu02} affords a powerful tool for a rigorous assignment of quantum ground states of geometrically frustrated Heisenberg spin systems at sufficiently high magnetic 
fields \cite{2zhit05,3derz06,krivnov2014,4derz15}.
Moreover, the localized nature of eigenstates plays an
important role in other flatband systems 
\cite{4derz15,mielke1992,mielke1993,6tasa98,7guzm12,bergholtz2013,9leyk13,VCM:PRL15,8mukh15,10stre17,leykam2018,11stre18,danieli_2021}.
This concept can be employed whenever destructive quantum  interference traps magnon(s) within a few lattice sites and hence, the frustrated quantum Heisenberg model can be exactly mapped onto a classical lattice-gas model with a hard-core potential \cite{2zhit05,3derz06,4derz15}. Using this approach, the microscopic nature of the last intermediate plateau emergent in a zero-temperature 
magnetization curve of the quantum spin-1/2 Heisenberg kagome antiferromagnet has been, for instance, elucidated along with the precise nature of a relevant second-order phase transition emerging 
at low but nonzero temperatures \cite{2zhit05,derzhko_bilay2010,PRL_schnack2020}. This exciting theoretical finding were 
experimentally verified by high-field 
magnetization measurement on kagome-like compound Cd-kapellasite CdCu$_3$(OH)$_6$(NO)$_3\cdot$H$_2$O \cite{okuma19}. The main advantage of the localized-magnon approach lies in that it also provides, besides an exact ground state, accurate description of
the low-temperature thermodynamics due to a proper counting of low-lying excited states \cite{2zhit05,3derz06,4derz15}. 

In our recent papers, we have provided a proper description of the low-temperature thermodynamics of the spin-1/2 
Heisenberg octahedral chain \cite{10stre17,11stre18} in the full range of the magnetic fields within the highly frustrated parameter region, because this frustrated quantum spin chain exhibits at sufficiently low magnetic fields another exact ground state with the character of the 
monomer-tetramer phase being composed from a localized two-magnon state. Exactly the same 
ground state with the character of monomer-tetramer phase appears in the ground-state phase 
diagram of the mixed spin-(1,1/2) Heisenberg octahedral chain, for which we have 
also found a consistent description of the low-temperature 
thermodynamics in the full range of the magnetic fields in a highly frustrated parameter region \cite{12karl19}. It should be mentioned that a less frustrated parameter region of the mixed spin-(1,1/2) Heisenberg octahedral chain involves besides the monomer-tetramer phase three additional fragmentized cluster-based Haldane phases, which are manifested in
the zero-temperature magnetization curve as fractional magnetization plateaus at 1/6, 1/9 and 1/12 of the saturation 
magnetization. The cluster-based Haldane phases appear due to a fragmentation of the octahedral chain, which is 
caused by a plaquette-singlet state incorporating four spins from an elementary square plaquette. Moreover, the 
plaquette-singlet state regularly appears within the cluster-based Haldane phases at certain periods of the octahedral chain and 
hence, smaller chain fragments separated from one another by the plaquette-singlet state can be alternatively 
considered as bound magnons effectively represented by hard-core dimers, trimers and tetramers, respectively. This 
fact gives us hope for a proper description of the low-temperature thermodynamics 
of the mixed spin-(1,1/2) Heisenberg 
octahedral chain in the full range of the magnetic fields also in a less frustrated parameter region from a mapping 
correspondence with the classical lattice-gas model of hard-core monomers and at least dimers. 

This paper is organized as follows. 
The model and basic steps of the calculation procedure are reviewed in Section \ref{model}. 
Section \ref{lgm} deals with the effective description of the model within the lattice-gas model of hard-core particles. 
Then in Section \ref{num_tools} we briefly illustrate our numerical tools,
namely the exact diagonalization (ED)
and the finite-temperature Lanczos method (FTLM). 
The most interesting results for the magnetization curves and thermodynamics are presented in Section \ref{results} 
together with numerical data obtained from ED  and FTLM.
Finally, some conclusions and future outlooks are mentioned in Section \ref{conclusion}. 

\section{Model and its ground states}
\label{model}
\begin{figure}
\begin{center}
\includegraphics[width=0.5\textwidth]{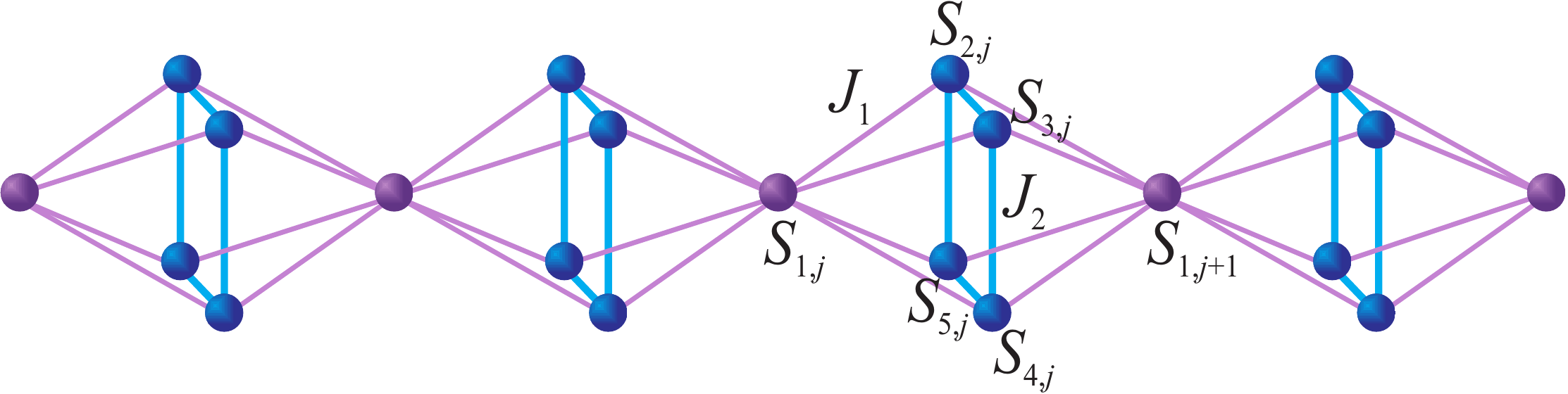}
\end{center}
\vspace{-0.5cm}
\caption{A schematic illustration of the mixed spin-(1,1/2) Heisenberg octahedral chain with spin-1 particles placed on monomeric sites and spin-1/2 particles on sites of square plaquettes. The coupling constant $J_1$ accounts for the antiferromagnetic interaction between nearest-neighbor monomeric and square-plaquette spins, while the interaction $J_2$ stands for the antiferromagnetic interaction between nearest-neighbor spins from square plaquettes.}
\label{figmod}       
\end{figure}

Let us consider the mixed spin-(1,1/2) Heisenberg octahedral chain, which is schematically illustrated in Fig. \ref{figmod} and given by the Hamiltonian
\begin{eqnarray}
\hat{\cal H} &=&
\sum_{j=1}^{N} \Bigl[ J_1 (\boldsymbol{\hat{S}}_{1,j} + \boldsymbol{\hat{S}}_{1,j+1}) \!\cdot\! (\boldsymbol{\hat{S}}_{2,j} + \boldsymbol{\hat{S}}_{3,j} + \boldsymbol{\hat{S}}_{4,j} + \boldsymbol{\hat{S}}_{5,j}) \Bigr.  \nonumber \\
&+& J_2 (\boldsymbol{\hat{S}}_{2,j}\!\cdot\!\boldsymbol{\hat{S}}_{3,j} + \boldsymbol{\hat{S}}_{3,j}\!\cdot\!\boldsymbol{\hat{S}}_{4,j}
+ \boldsymbol{\hat{S}}_{4,j}\!\cdot\!\boldsymbol{\hat{S}}_{5,j} + \boldsymbol{\hat{S}}_{5,j}\!\cdot\!\boldsymbol{\hat{S}}_{2,j})   \Bigr. \nonumber \\
&-&\Bigl.h \sum_{i=1}^{5}\hat{S}_{i,j}^{z} \Bigr],
\label{hamoskamos}
\end{eqnarray}
where $\boldsymbol{\hat{S}}_{i,j} \equiv (\hat{S}_{i,j}^x, \hat{S}_{i,j}^y, \hat{S}_{i,j}^z)$ denotes spatial components of the spin-1 (spin-1/2) operator for the subscript $i=1$ ($i=2,3,4,5$). The exchange interaction $J_1>0$ accounts for the antiferromagnetic Heisenberg  interaction between monomeric spins and spins placed in vertices of square plaquette, while the spins belonging to the same square plaquette are coupled through the antiferromagnetic exchange interaction $J_2>0$. The last term in the Hamiltonian (\ref{hamoskamos}) 
accounts for the Zeeman energy of magnetic moments in the external magnetic field $h \geq 0$ and $N$ marks the total number of five-spin unit cells of the octahedral chain.
Periodic boundary conditions are assumed  $\boldsymbol{S}_{1,N+1} \equiv \boldsymbol{S}_{1,1}$ in order to eliminate boundary effects.

Before proceeding to a discussion of the calculation method to be used for thermodynamic description, let us briefly recall all available ground states of the mixed-spin Heisenberg octahedral chain, which were comprehensively studied in our previous paper \cite{12karl19}. 
\begin{figure}
\begin{center}
\includegraphics[width=0.5\textwidth]{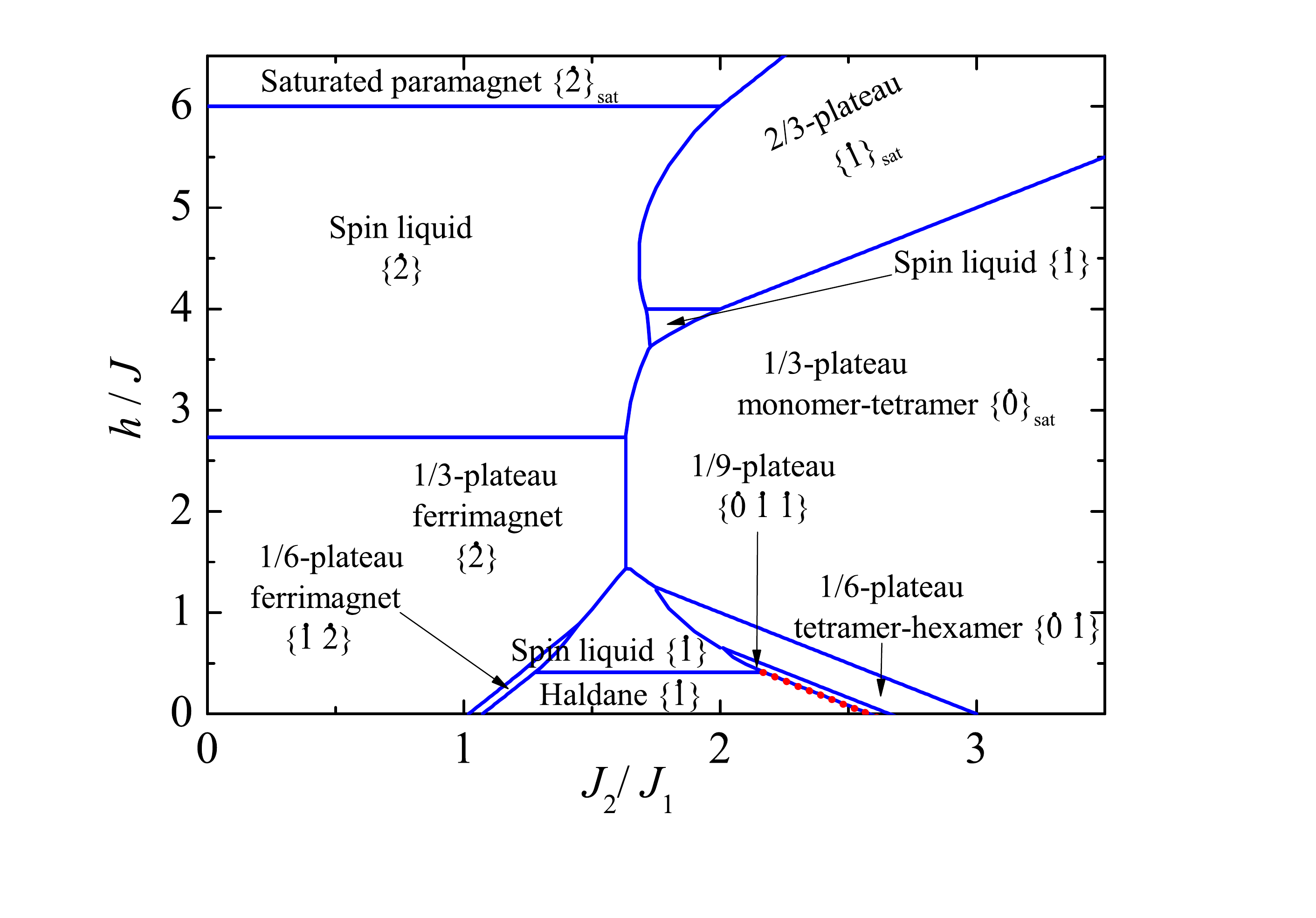}
\end{center}
\vspace{-1.3cm}
\caption{The ground-state phase diagram of the mixed spin-(1,1/2) Heisenberg octahedral chain. The numbers in curly brackets denote the value of composite spins of square plaquettes (all phases with more than one number 
correspond to a phase with spontaneously broken symmetry). The tiny phase illustrated by red dots corresponds to the 1/12-plateau in the zero-temperature magnetization curve, which has  each fourth square plaquette in a singlet state and other three consecutive square plaquettes in a triplet state, i.e.  \{$\dot{0}\dot{1}\dot{1}\dot{1}$\}.}
\label{GSPD}       
\end{figure}
The ground-state phase diagram shown in Fig. \ref{GSPD} was obtained in our previous work by the use of a few complementary analytical and numerical methods \cite{12karl19}. As one can see from Fig. \ref{GSPD}, there are two ferrimagnetic phases corresponding to 1/6- and 1/3-plateaus, two kinds of spin-liquid phases, four fragmentized cluster-based Haldane phases corresponding to the 1/3-, 1/6-, 1/9- and 1/12-plateaus,
a bound-magnon crystal phase corresponding to 2/3-plateau as well as the Haldane phase. In our recent work \cite{12karl19} we 
have investigated thermodynamics of the mixed-spin Heisenberg octahedral chain in the highly frustrated parameter region $J_2/J_1>3$ by using of the mapping correspondence with the classical lattice-gas model, which takes into consideration bound two-magnon and one-magnon states of square plaquettes as two different monomeric particles of the effective lattice-gas model of hard-core monomers. More concretely, the highly frustrated parameter region includes besides the 
trivial fully polarized ferromagnetic state also 
the bound magnon-crystal phase schematically shown in Fig. \ref{figfazy}(a) and given by the eigenvector 
\begin{eqnarray}
\!\!\!\!\!|{\rm BM}\rangle = \prod_{j=1}^N \! |1\rangle_{1,j} \!\otimes\! \frac{1}{2}
(|\!\!\downarrow_{2,j}\uparrow_{3,j}\uparrow_{4,j}\uparrow_{5,j}\rangle 
-|\!\!\uparrow_{2,j}\downarrow_{3,j}\uparrow_{4,j}\uparrow_{5,j}\rangle \nonumber \\
+|\!\!\uparrow_{2,j}\uparrow_{3,j}\downarrow_{4,j}\uparrow_{5,j}\rangle
-|\!\!\uparrow_{2,j}\uparrow_{3,j}\uparrow_{4,j}\downarrow_{5,j}\rangle\!)\!  \nonumber\\
\label{BM}
\end{eqnarray}
and the monomer-tetramer phase shown in Fig. \ref{figfazy}(b) and given by the eigenvector 
\begin{eqnarray}
|{\rm MT} \rangle = \prod_{j=1}^N \! |S\rangle_{1,j} \!\otimes\! 
\Bigl[\!\frac{1}{\sqrt{3}}(|\!\!\uparrow_{2,j}\downarrow_{3,j}\uparrow_{4,j}\downarrow_{5,j}\rangle   \!\!+\!\! |\!\!\downarrow_{2,j}\uparrow_{3,j}\downarrow_{4,j}\uparrow_{5,j}\rangle)  \nonumber \\
- \frac{1}{\sqrt{12}} (|\!\!\uparrow_{2,j}\uparrow_{3,j}\downarrow_{4,j}\downarrow_{5,j}\rangle  
+   |\!\!\uparrow_{2,j}\downarrow_{3,j}\downarrow_{4,j}\uparrow_{5,j}\rangle \nonumber \\   
 +  |\!\!\downarrow_{2,j}\uparrow_{3,j}\uparrow_{4,j}\downarrow_{5,j}\rangle + |\!\!\downarrow_{2,j}\downarrow_{3,j}\uparrow_{4,j}\uparrow_{5,j}\rangle) \Bigr]\!, \nonumber \\
\label{MT}
\end{eqnarray}
where $|S\rangle_{1,j}$ denotes one out of three available states $|\pm 1\rangle_{i,j}$ and $|0\rangle_{i,j}$ of the monomeric spins.
In the following we will extend the calculation procedure in order to account for another two ground states, the tetramer-hexamer phase and Haldane phase, which are schematically shown in Fig. \ref{figfazy}(c) and (d).  

\begin{figure}[h]
\begin{center}
\includegraphics[width=0.5\textwidth]{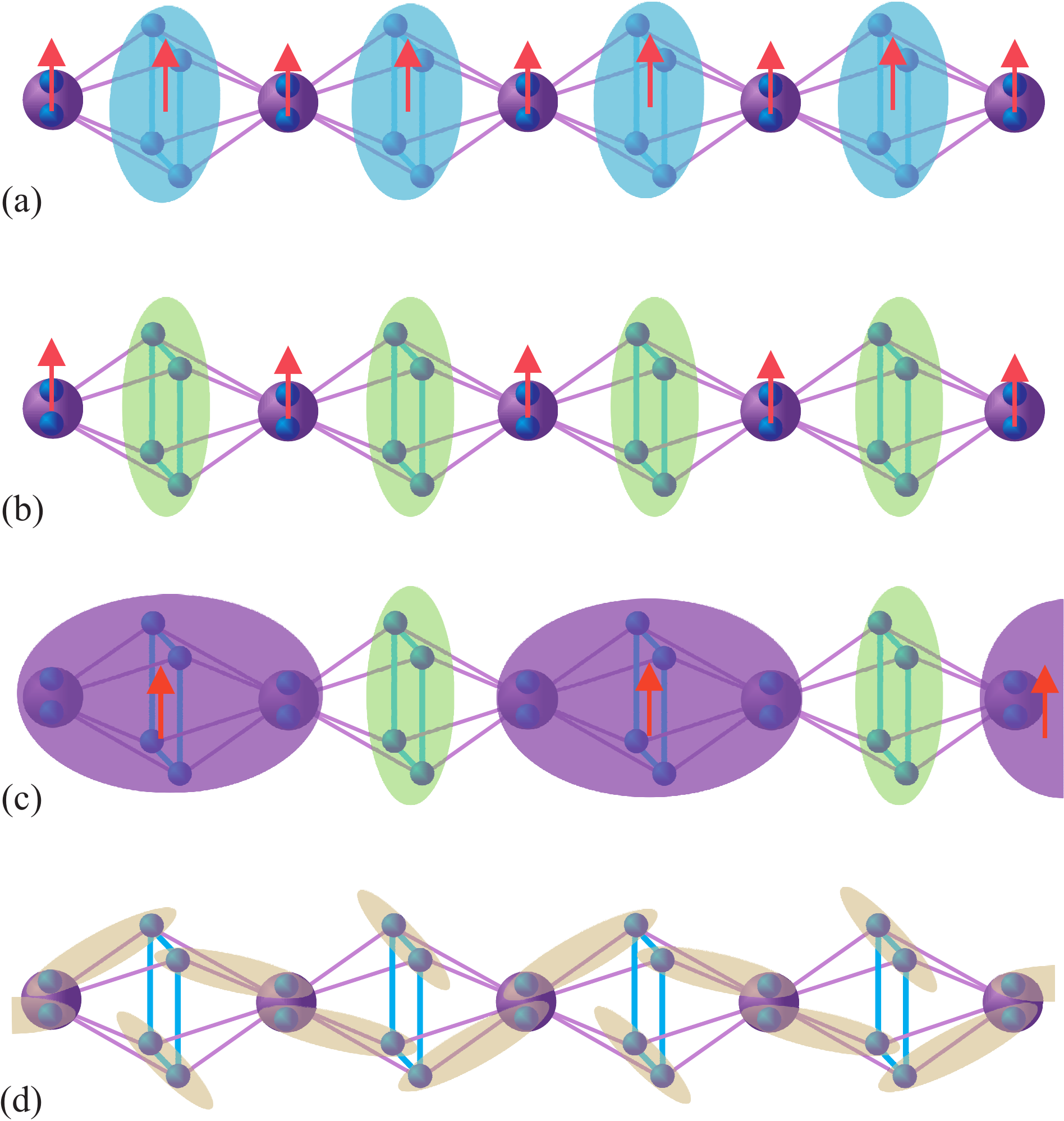}
\end{center}
\vspace{-0.5cm}
\caption{A schematic illustration of a few selected ground states of the mixed-spin Heisenberg octahedral chain given by the 
Hamiltonian (\ref{hamoskamos}), which are realized in the ground-state phase diagram in the moderately frustrated regime: (a) the bound magnon-crystal phase; (b) the monomer-tetramer phase; (c) the tetramer-hexamer phase; (d) the Haldane phase. Red arrows represent polarized triplet state of a given cluster shaded by the different colors, whereby spin clusters without arrows are in a singlet state.}
\label{figfazy}       
\end{figure}

\section{Effective lattice-gas model}
\label{lgm}
In order to obtain magnetothermodynamics, we will use the mapping correspondence between the Hamiltonian (\ref{hamoskamos}) of the mixed-spin Heisenberg octahedral chain  and the corresponding lattice-gas model. For the highly frustrated parameter region $J_2/J_1>3$ we have presented thermodynamic properties in our previous paper \cite{12karl19}.  The Hamiltonian (\ref{hamoskamos}) can be in the moderately frustrated regime $J_2/J_1 \in (2.16; 3)$ mapped to the effective monomer-dimer lattice-gas model given by the Hamiltonian 
\begin{eqnarray}
H_{\rm eff}&=&E_{\rm FM}^0-3Nh-\mu_H^{(0)}n_H-\mu_1^{(2)}\sum_{j=1}^Nm_j \nonumber \\
&-&\mu_1^{(1)}\sum_{j=1}^Nn_j-\sum_{n=3}^5\mu_{2}^{(n)}\sum_{j=1}^Nd_j,
\label{hef}
\end{eqnarray}
which can be developed by the following construction. From the energy of the 
fully polarized ferromagnetic state $E_{\rm FM}^0=N(4J_1+J_2)$ in zero field we have subtracted the energies of two kinds of monomeric particles whose presence or absence is determined by occupation numbers $m_j$ and $n_j$ and a dimeric particle determined through the occupation number $d_j$. The chemical potential of the first monomeric particle $\mu_1^{(1)}=2J_1 +2J_2-h$ determines an energy cost, which is associated with creation of bound one-magnon eigenstate on a square plaquette on a fully polarized ferromagnetic background, while the chemical potential of the second monomeric particle $\mu_2^{(2)}=4J_1 +3J_2-2h$ determines an energy cost connected with creation of the singlet-tetramer state on the ferromagnetic background. The chemical potential of the dimeric particle $\mu_2^{(n)} = 7J_1+2J_2-nh \quad (n=3,4,5)$ represents an energy cost, which relates 
to the creation of a single octahedron in one out of three available  lowest-energy triplet states.  All possible spin values of the triplet state $S_T^z=\pm 1, 0$, were taken into consideration in order to obtain correct degeneracy of the triplet-hexamer state in a zero-field limit. The gapped Haldane phase has been introduced as a particle spread over the whole chain as given by the occupation number $n_H$.  The chemical potential $\mu_H^{(0)}=2NJ_2+2NJ_1(2-\epsilon_N)$ determines the energy difference between the Haldane phase and the fully  polarized ferromagnetic state, where $\epsilon_N$ denotes ground-state energy of the Haldane phase per spin for a spin-1 Heisenberg chain of $N$ sites with unit coupling constant. The energy gap of the Haldane phase is well known $\epsilon_{\infty}\approx -1.401484$ in the thermodynamic limit $N \to \infty$, while finite-size results for the chain lengths of eight and twelve spins corresponding to four and six unit cells of octahedral chain equal to $\epsilon_4=-1.417119$ and $\epsilon_6=-1.405796$, respectively.
 
\begin{figure}
\begin{center}
\includegraphics[width=0.5\textwidth]{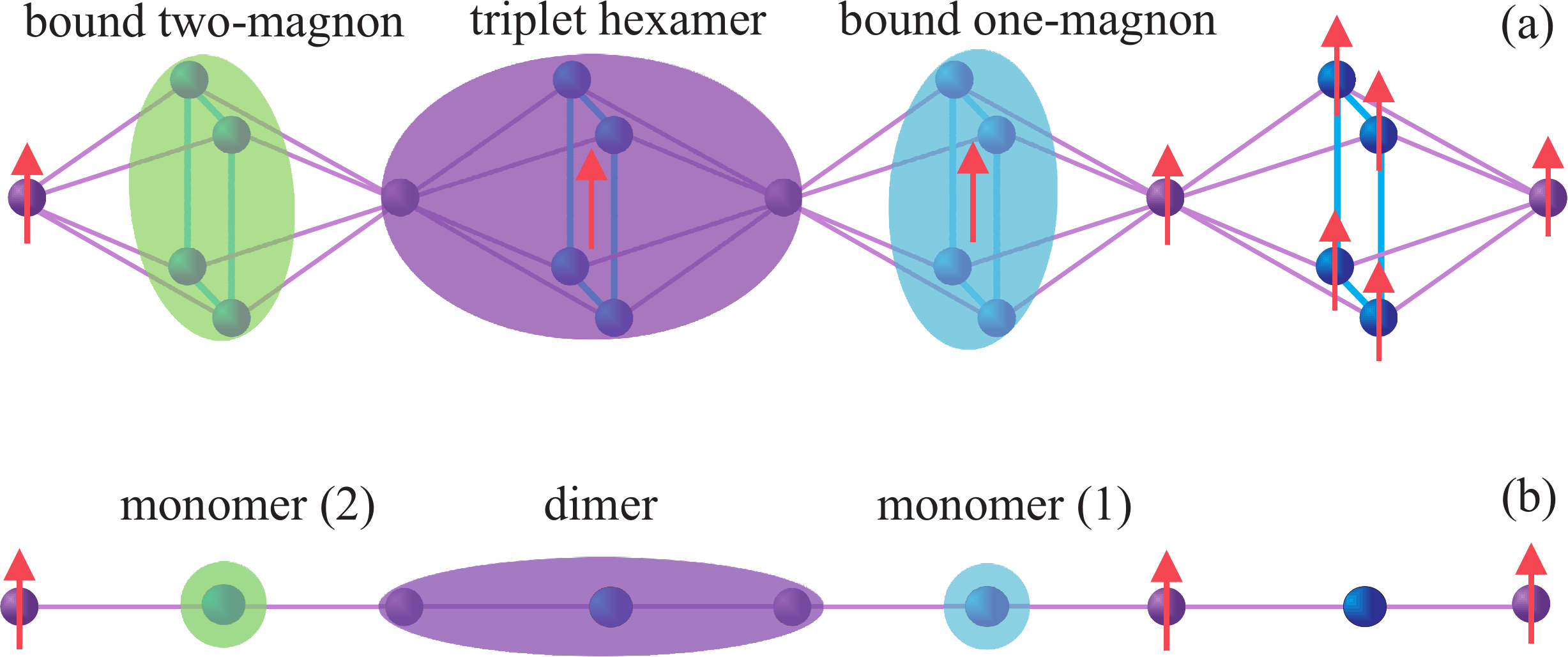}
\end{center}
\vspace{-0.5cm}
\caption{A schematic illustration of the mixed-spin Heisenberg octahedral chain, in which spins from $i$-th elementary unit are either in  bound one-magnon or two-magnon states, triplet-hexamer state or are fully polarized. Lower panel represents a schematic illustration of the corresponding lattice-gas model with two kinds of monomeric particles and dimeric particles.}
\label{figlg}       
\end{figure}

Except the trivial fully polarized ferromagnetic state the effective Hamiltonian (\ref{hef}) takes into account a few lowest-energy states such as bound one-magnon and two-magnon states of square plaquettes as well as a triplet state of an elementary octahedron composed from a single square plaquette and its two neighboring monomeric spins (see Fig. \ref{figlg}). In addition, the effective Hamiltonian (\ref{hef}) also takes into account the Haldane phase spread over the whole octahedral chain.  
\label{latgas}
The partition function corresponding to the effective Hamiltonian (\ref{hef}) reads
\begin{eqnarray}
{\cal {Z}}&=&\exp(-\beta E_{\rm FM}^0+3\beta Nh)\sum_{\{m_j\}}\sum_{\{n_j\}}\sum_{\{d_j\}}\sum_{n_H^0=1,0}\nonumber \\
&\times&\prod_{j=1}^N(1-d_jd_{j+1})(1-m_jd_j)(1-n_jd_j)(1-m_jn_j) \nonumber \\
&\times&(1-n_Hn_j)(1-n_Hm_j)(1-n_Hd_j)\nonumber \\
&\!\!\!\times&\sum_{n=3}^5\!\exp\!\!\left(\beta \mu_H^{(0)}n_h\!+\!\beta\mu_1^{(2)}m_j\!+\!\beta\mu_1^{(1)}n_j\!+\!\beta\mu_2^{(n)}d_j\right)\!\!,
\end{eqnarray}
where $\beta=1/(k_{\rm B}T)$,  $k_{\rm B}$ is the Boltzmann's constant, $T$ is absolute temperature and the projection operators $(1-d_jd_{j+1})(1-m_jd_j)(1-n_jd_j)(1-m_jn_j)(1-n_Hn_j)(1-n_Hm_j)(1-n_Hd_j)$ forbids the multiple occupancy of the square plaquette with more than one  particle of the effective lattice-gas model.  After tracing out degrees of freedom of the monomeric particles $m_j$ and $n_j$, as well as, of the Haldane phase $n_H$, the problem of finding the partition function reduces to a problem of finding all eigenvalues of the transfer matrix
\begin{eqnarray}
{\cal {Z}}&=&\exp(-\beta E_H\!)\!+\!\exp(-\beta E_{\rm FM}^0\!\!+\!3\beta Nh)\!\!\sum_{\{d_j\}}\!\prod_{j=1}^N\!{\rm T}(d_j, \!d_{j\!+\!1}) \nonumber \\
&=&\exp(-\beta E_H\!)\!+\!\exp(-\beta E_{\rm FM}^0\!\!+\!3\beta Nh){\rm Tr~T}^N,
\end{eqnarray}
where $E_H=-NJ_2+2NJ_1\epsilon_N$ is the energy of Haldane phase. The expression ${\rm T}(d_j, \!d_{j\!+\!1})$ denotes the transfer matrix depending only on occupation numbers of the dimeric particles from two adjacent lattice sites satisfying the hard-core constraint 
\begin{eqnarray}
{\rm T}(d_j, \!d_{j+1})\!\!&=&\!\! (1-d_jd_{j+1})\sum_{n=3}^5\exp(\beta\mu_2^{(n)}d_j)  \nonumber \\
\!\!&\times&\!\!\left[1+(1-d_j)\exp\left(\beta\mu_1^{(2)}+\beta\mu_1^{(1)}\right)\right].
\label{tm}
\end{eqnarray}
 The transfer matrix as defined by Eq. (\ref{tm}) has the following matrix representation
 \begin{eqnarray}
\!\!\!\!\!\!&&{\rm T}(d_j, \!d_{j+1})= \nonumber \\
\!\!\!\!\!\!&&\!\!\!\!\!\!\left(\!\!\!\begin{array}{cc}
1\!+\!\exp(\beta\mu_1^{(1)})\!+\!\exp(\beta\mu_1^{(2)}) & 1\!+\!\exp(\beta\mu_1^{(1)})\!+\!\exp(\beta\mu_1^{(2)})\\
\exp(\beta\mu_2^{(4)})[1\!\!+\!\!2\cosh(\beta h)]& 0 
\end{array}
\!\!\!\right).\nonumber \\
\label{rov8}
\end{eqnarray}
After the diagonalization of the transfer matrix (\ref{rov8}), one gets two eigenvalues
\begin{eqnarray}
\lambda_{\pm}\!=\!\frac{1}{2}\left(\Xi\!\pm\!\sqrt{\Xi^2\!+\!4\Xi\exp(\beta\mu_2^{(4)})[1\!+\!2\cosh(\beta h)]}\right)\!\!,
\end{eqnarray}
where $\Xi=1+\exp(\beta \mu_1^{(1)})+\exp(\beta\mu_1^{(2)})$. Then, the partition function of the monomer-dimer lattice-gas model (\ref{hef}) is given by the equation
\begin{eqnarray}
{\cal Z}\!\!=\!\exp(-\beta E_H)\!+\!\exp(-\beta E_{\rm FM}^0\!+\!3\beta Nh)\!\!\left(\lambda_+^N\!\!+\!\lambda_-^N\right)\!\!.
\end{eqnarray}
In the thermodynamic limit $N \to \infty$ the partition function is given only by the higher eigenvalue of the transfer matrix
\begin{eqnarray}
{\cal Z}_{\infty}=\exp(-\beta E_H)+\exp(-\beta E_{\rm FM}^0+3\beta Nh)\lambda_+^N. \nonumber \\
\end{eqnarray}
From the partition function one can obtain the free energy, as well as the magnetization,
the susceptibility, the entropy and the specific heat. 
 \begin{figure*}
\begin{center}
\includegraphics[width=0.5\textwidth]{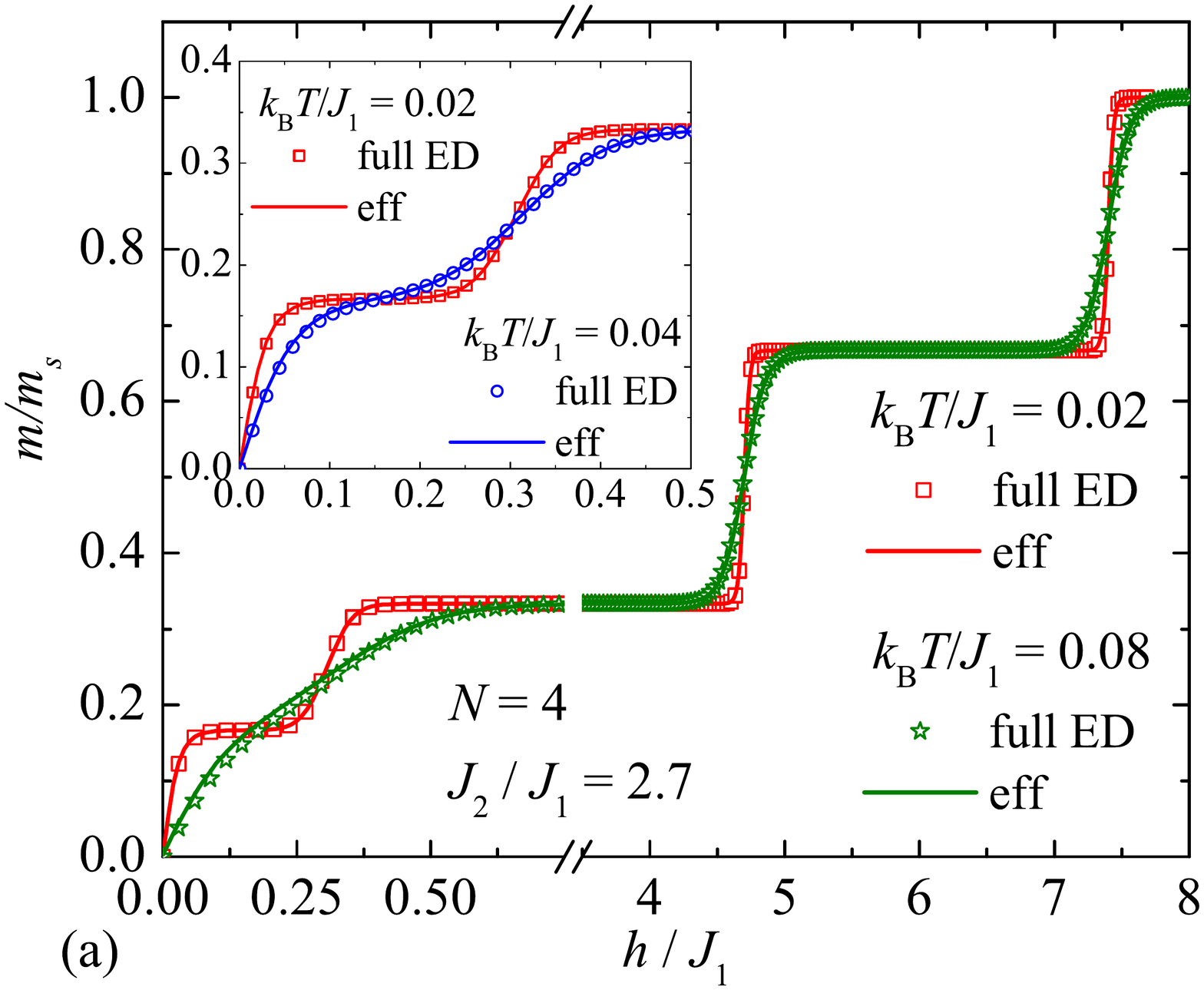}
\hspace*{-1cm}
\includegraphics[width=0.5\textwidth]{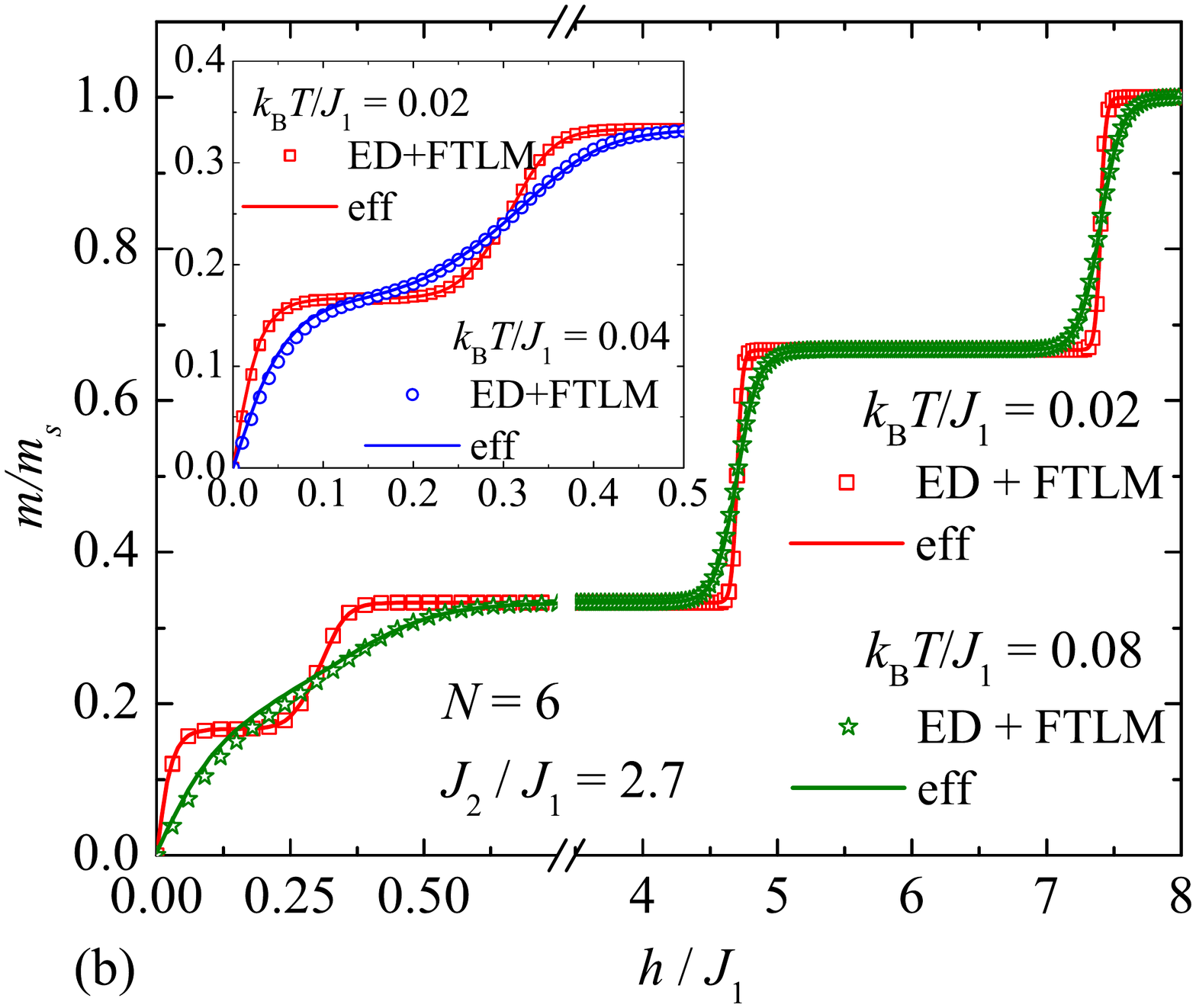}
\includegraphics[width=0.5\textwidth]{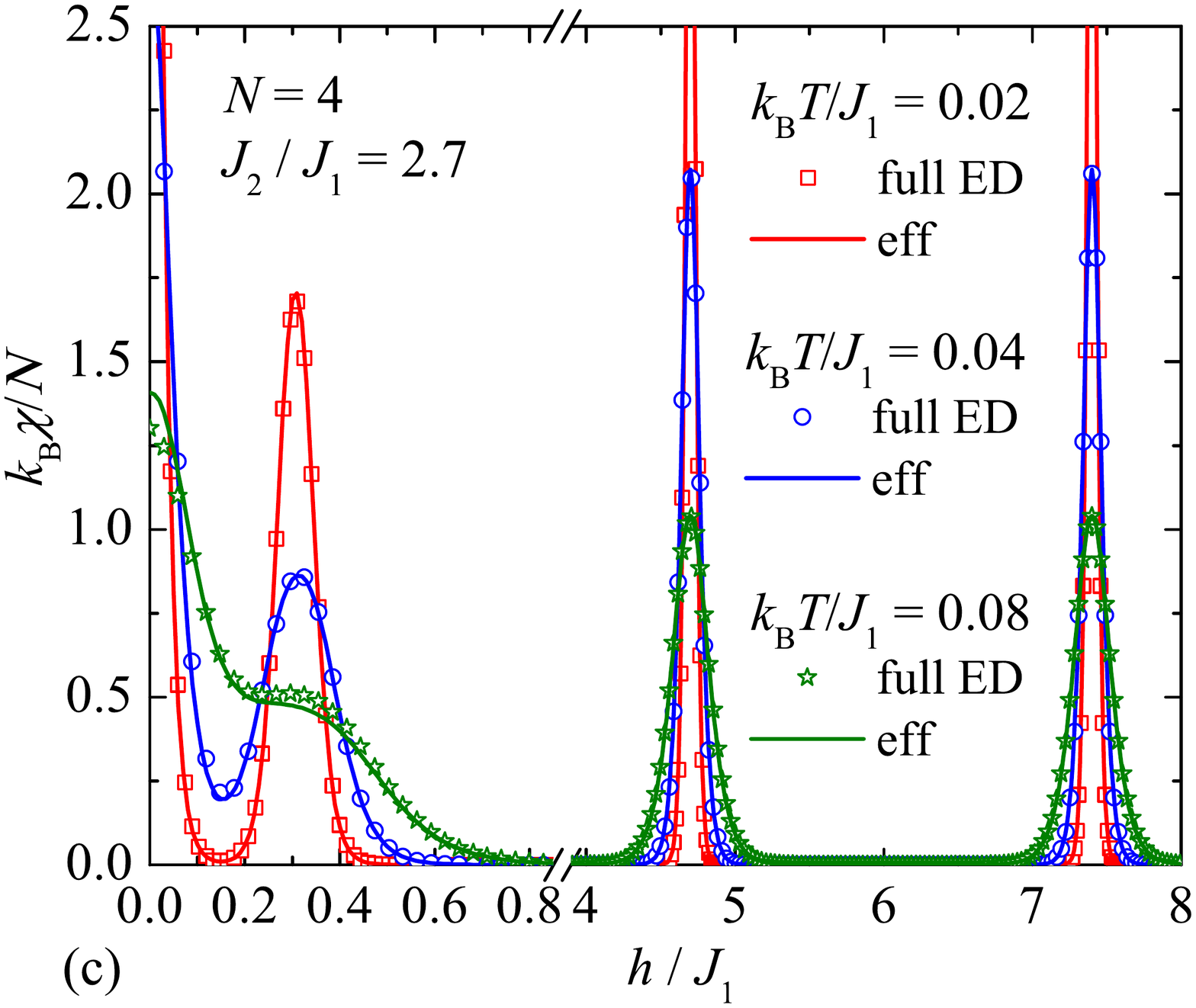}
\hspace*{-1cm}
\includegraphics[width=0.5\textwidth]{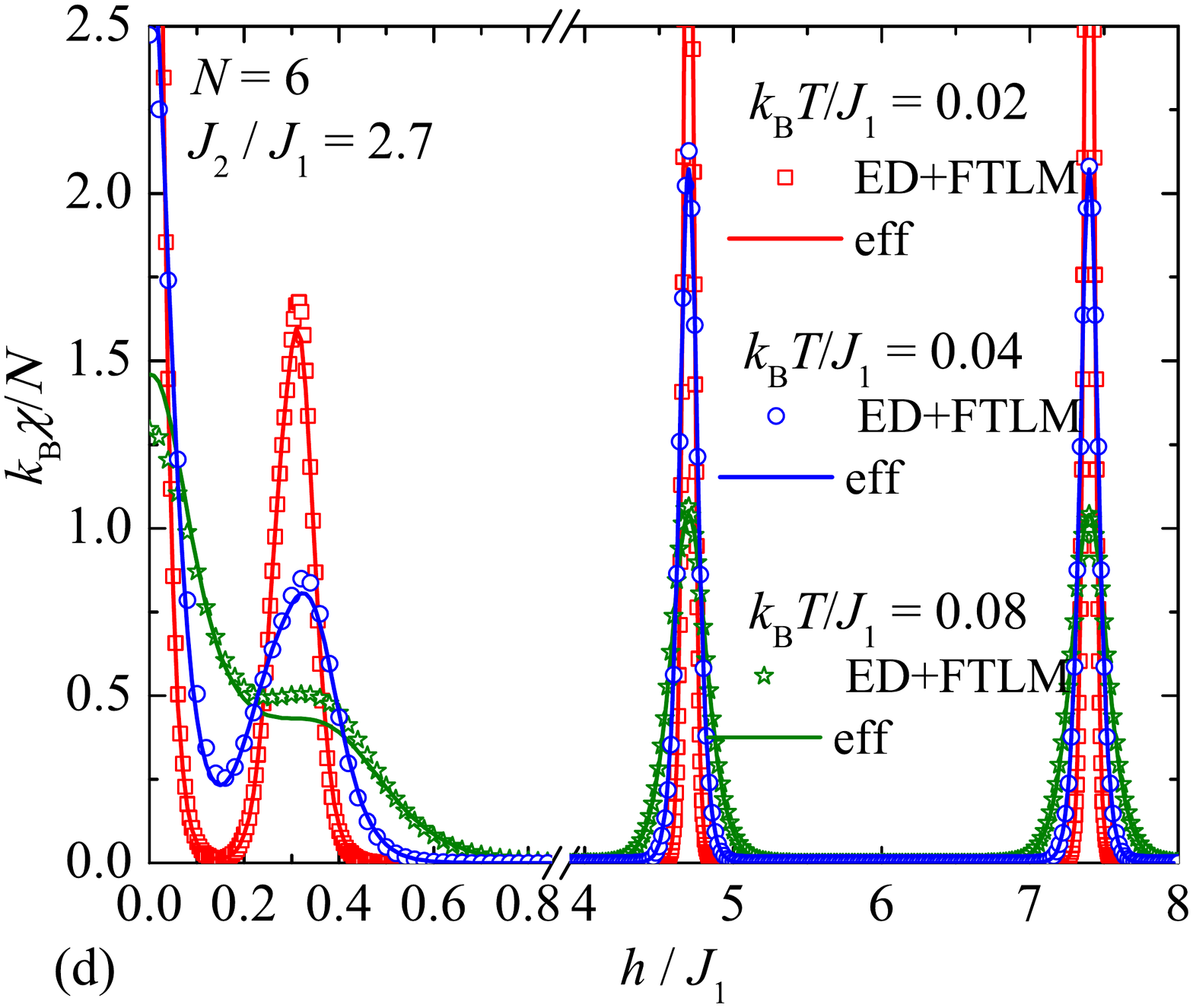}
\end{center}
\vspace{-1cm}
\caption{A comparison of ED and FTLM data shown by symbols with the analytical results obtained from the effective lattice-gas model shown by solid lines for magnetic-field dependencies of (a)-(b) magnetization and (c)-(d) susceptibility of the mixed-spin Heisenberg octahedral chain with $N=4$ (left panel) and $N=6$ (right panel) unit cells, the fixed value of the interaction ratio $J_2/J_1=2.7$ and a few selected values of temperature.
The magnetization is normalized with respect to its saturation value and susceptibility is normalized per unit cell.}
\label{fig1}       
\end{figure*}

\section{Exact-diagonalization and finite-temperature Lanczos methods }
\label{num_tools}

In order to verify the results obtained from the effective lattice-gas model, 
we use the  full exact numerical diagonalization (ED) and the finite temperature Lanczos method
(FTLM). 
Trivially  the Hamiltonian (\ref{hamoskamos}) commutes with
the $z$-component of the total spin $\hat{S}_T^z$, i.e., the Hilbert space
splits into orthogonal subspaces related to the eigenvalues $S_T^z$ of
$\hat{S}_T^z$.
In addition, 
we also   
exploit lattice symmetries to further split the $S_T^z$ subspaces into smaller
symmetry related subspaces. 
For that we use J\"org
Schulenburg's {\it spinpack}
code
\cite{49,50}  to perform the ED and the FTLM calculations.

The ED is a well established quantum many-body technique which is
widely applied to frustrated quantum spin systems, see, e.g.,
Ref.~\cite{lauchli}. 
The FTLM is an unbiased accurate numerical approximation by which 
the partition function ${\cal {Z}}$ is determined using trace estimators
\cite{51,52,53,54,55,56,57,58,59,60}.
${\cal {Z}}$ is then given by  a Monte-Carlo like representation, 
i.e., the sum over a complete set of basis states entering ${\cal {Z}}$ 
is replaced by a much smaller sum over $R$ random vectors $|{\nu}\rangle$ for each
symmetry-related orthogonal subspace 
${\mathcal H}(M,\gamma)$ of the Hilbert space, where $\gamma$
labels the irreducible representations of the employed
symmetries. 

In the present case we use full ED to calculate the
partition function  ${\cal {Z}}$ for
chains of $N=4$ unit cells (i.e. 20 sites).   
For a longer  chain of $N=6$ unit cells (i.e. 30 sites) we combine both methods.
We use the full ED to determine  the contribution to
${\cal {Z}}$ of the  upper sectors of
$|S_T^z|=18,\ldots,12$ and the FTLM to
calculate  the contribution of the lower sectors of
$|S_T^z| \le 11$ to ${\cal {Z}}$.

\section{Results and discussion}
\label{results}

\begin{figure*}
\begin{center}
\vspace{-1cm}
\includegraphics[width=0.5\textwidth]{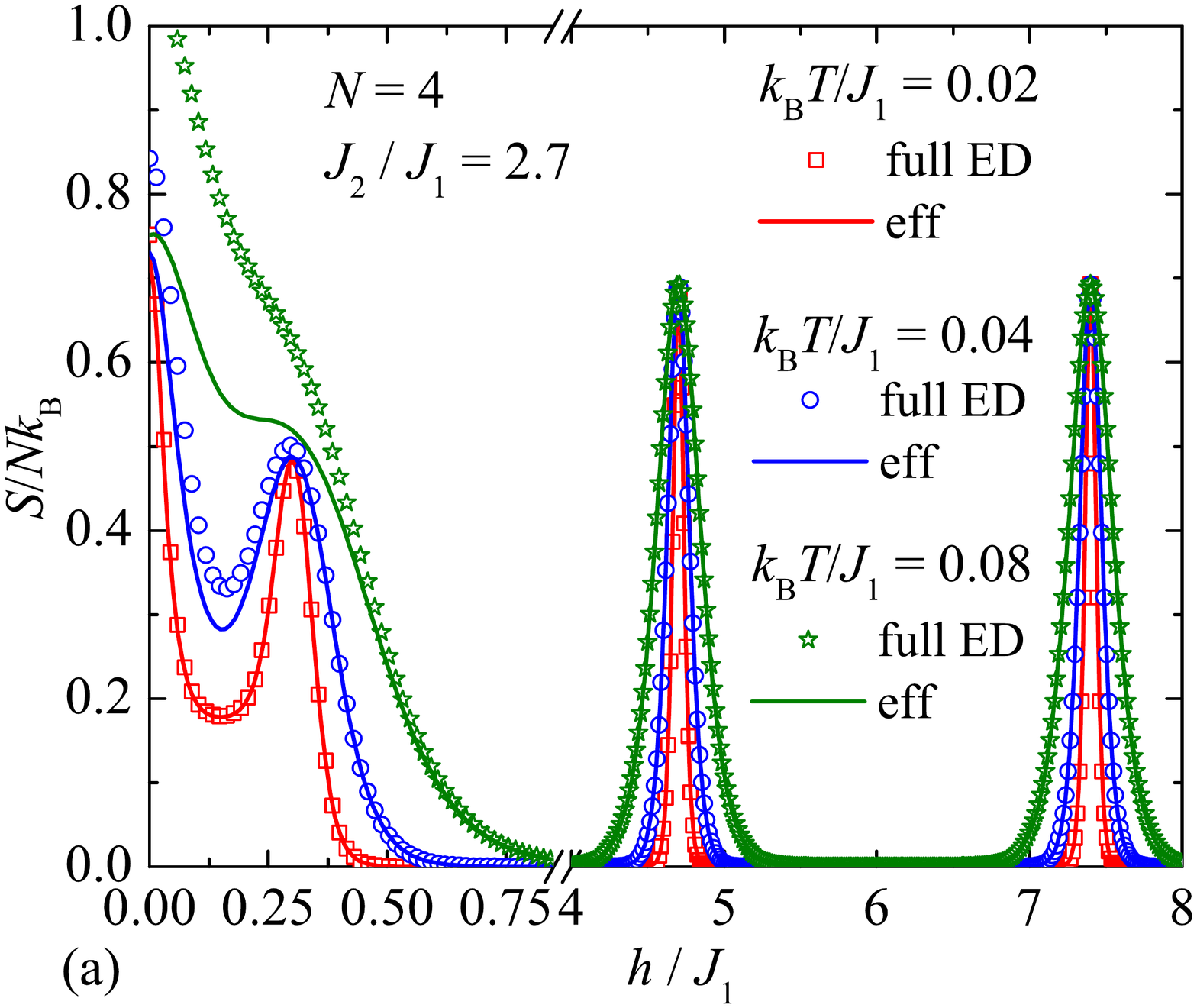}
\hspace{-1cm}
\includegraphics[width=0.5\textwidth]{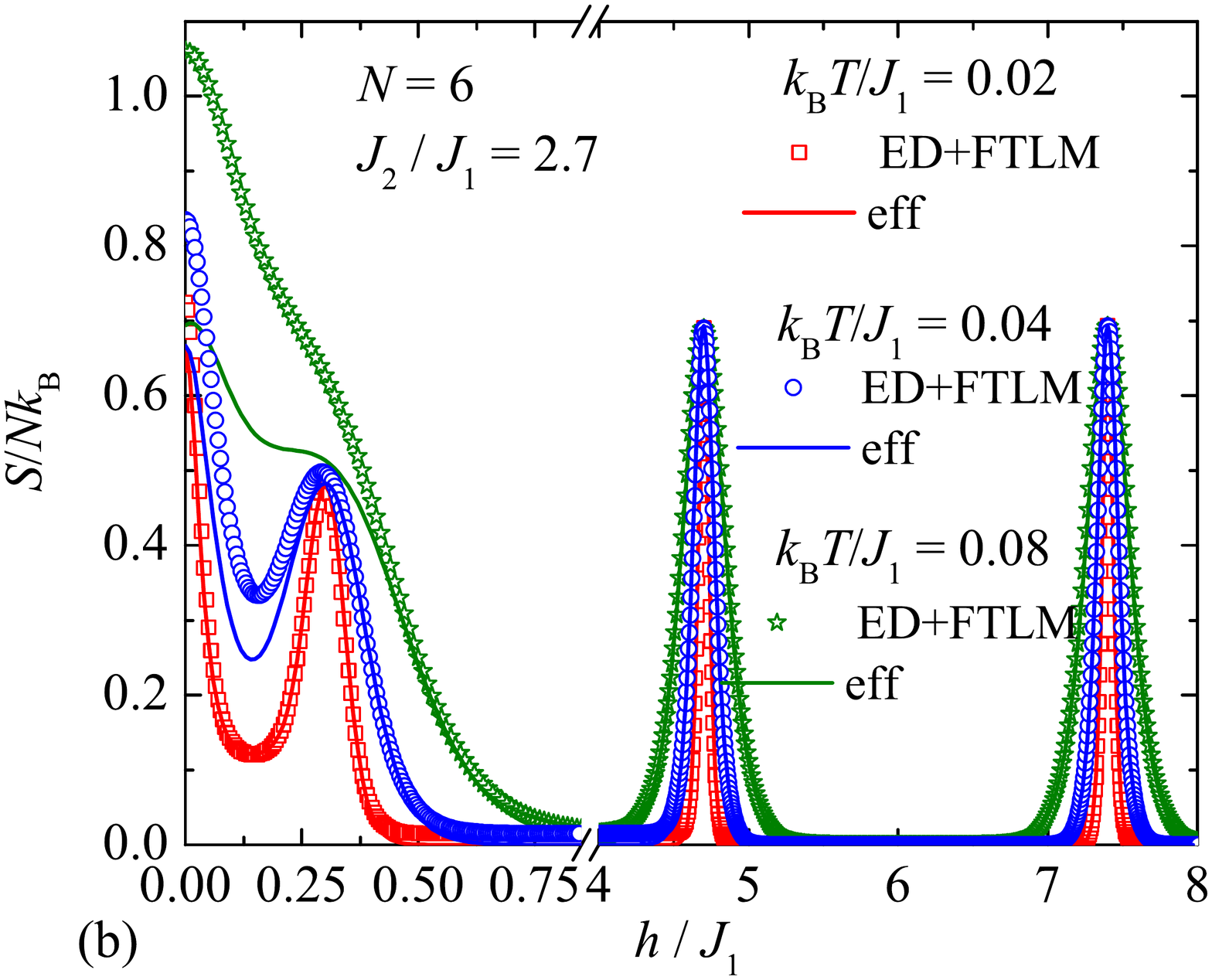}
\vspace{-1cm}
\includegraphics[width=0.5\textwidth]{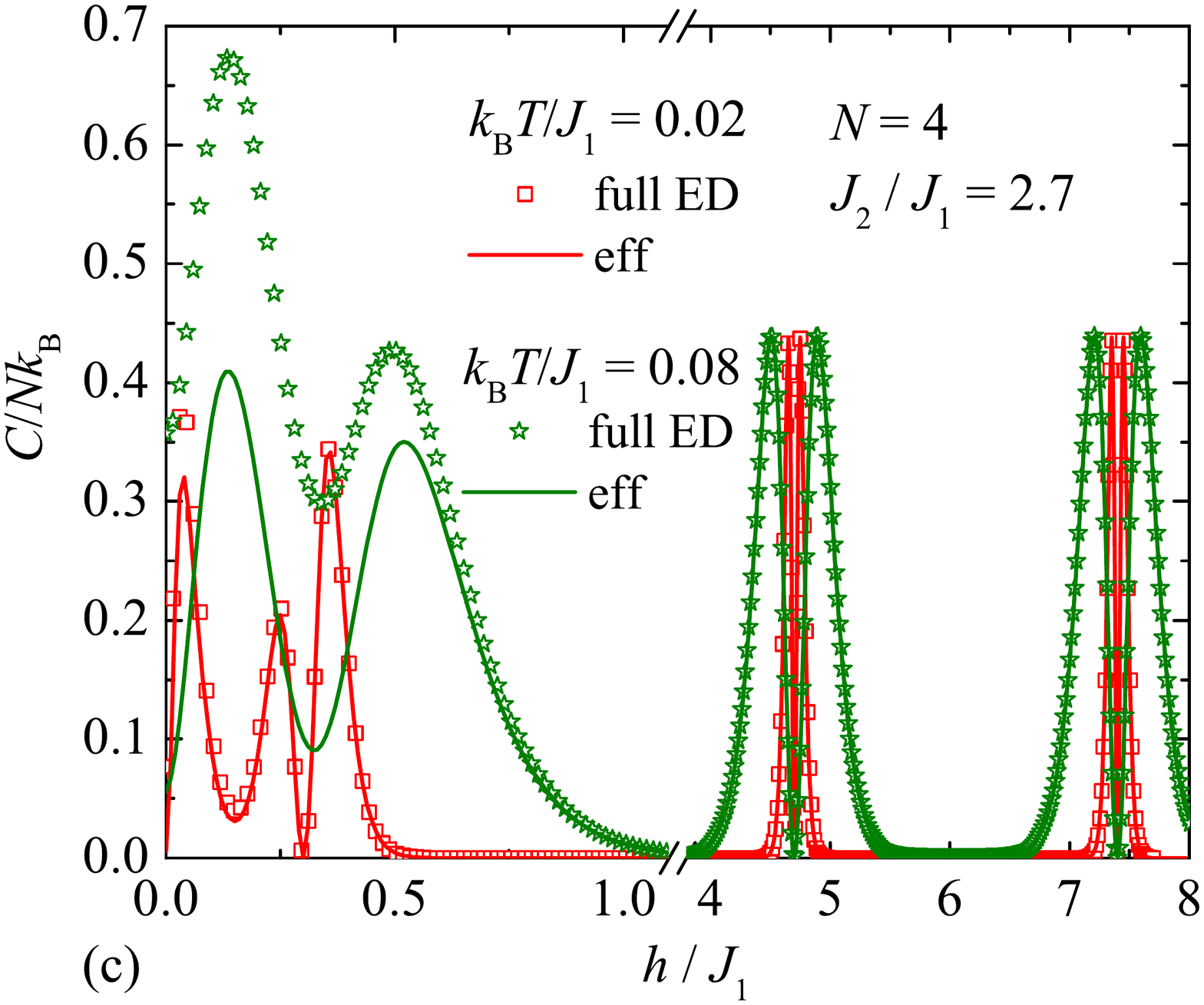}
\hspace{-1cm}
\includegraphics[width=0.5\textwidth]{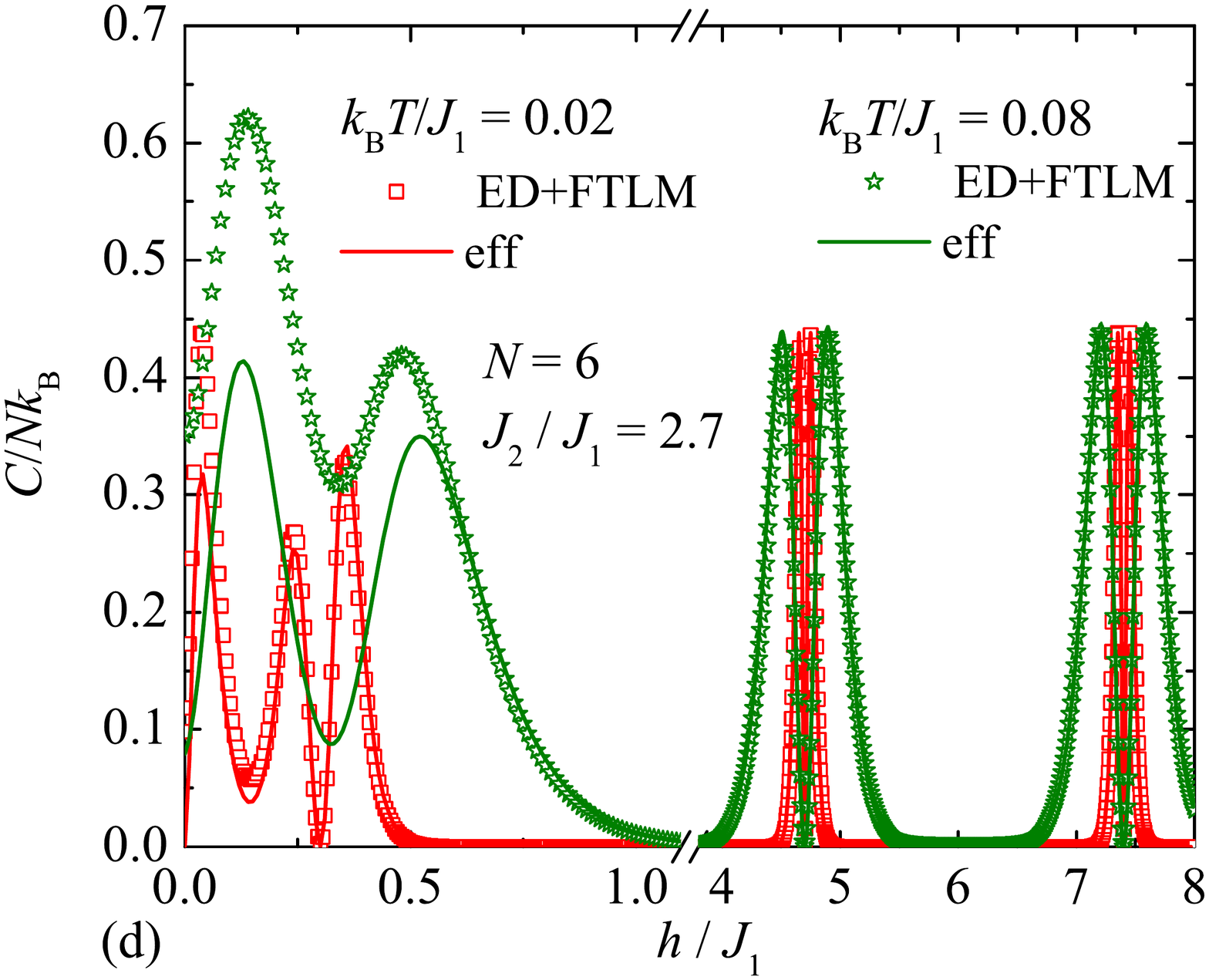}
\end{center}
\caption{A comparison of ED and FTLM data shown by symbols with the analytical results obtained from the effective lattice-gas model shown by solid lines for magnetic-field dependencies of (a)-(b) entropy and (c)-(d) specific heat of the mixed-spin Heisenberg octahedral chain with $N=4$ (left panel) and $N=6$ (right panel) unit cells, the fixed value of the interaction ratio $J_2/J_1=2.7$ and a few selected values of temperature. Both quantities are normalized per unit cell.}
\label{fig2}       
\end{figure*}

Let us proceed to a discussion of the most interesting results. The magnetization curves of the mixed spin-(1, 1/2) Heisenberg octahedral chain for four and six elementary unit cells $N=4$ and $N=6$ (i.e. 20 spins and 30 spins in total, respectively) are plotted in Figs. \ref{fig1}(a)-(b) for the fixed value of the interaction ratio $J_2/J_1=2.7$. In accordance with the ground-state phase diagram, one can find in the magnetization process of the mixed spin-(1, 1/2) Heisenberg octahedral chain three intermediate magnetization plateaus at 1/6, 1/3 and 2/3 of its saturation magnetization. It directly follows from a comparison of full ED and FTLM (both shown by symbols) with the results obtained from the effective model (solid lines) that the analytical results from the simplified lattice-gas model are in perfect agreement with 
precise numerical results in the full range of magnetic fields up to the temperature $k_{\rm B}T/J_1\lesssim 0.08$.
The insets in Fig. \ref{fig1}(a) and (b) are focused on magnetization curves calculated in
the low-field region for two low enough temperatures $k_{\rm B}T/J_1=0.02$ and 0.04, where a tiny 1/6-plateau due to the cluster-based Haldane phase with the 
character of a tetramer-hexamer state is present.
The presence of the 1/6-plateau is satisfactorily described by the effective lattice-gas model including hard-core monomeric and dimeric particles, the latter of which are crucial for its correct description.

 A perfect agreement of ED and FTLM data with the analytical results obtained from the effective lattice-gas model is present also in the magnetic-field dependence of the susceptibility depicted in Fig. \ref{fig1}(c) and \ref{fig1}(d) for $N=4$ and
$N=6$ elementary unit cells, the interaction ratio $J_2/J_1=2.7$ and three selected values of temperature. The susceptibility as a function of magnetic field exhibits peaks around each field-driven phase transition, which become lower and rounder upon increasing of the temperature. Besides the perfect agreement of numerical and analytical calculations, there is only
a small visible deviation of analytical calculations from the full ED data located in the magnetic-field range $h/J_1\in(0.2, 0.4)$ at temperature $k_{\rm B}T/J_1=0.08$ for four unit cells [Fig. \ref{fig1}(c)], whereas
a somewhat greater discrepancy observable already for the same field range at lower 
temperature $k_{\rm B}T/J_1=0.04$ for six unit cells [Fig. \ref{fig1}(d)]
might be attributed to the approximate nature of FTLM data.

\begin{figure*}
\begin{center}
\vspace{-1cm}
\includegraphics[width=0.5\textwidth]{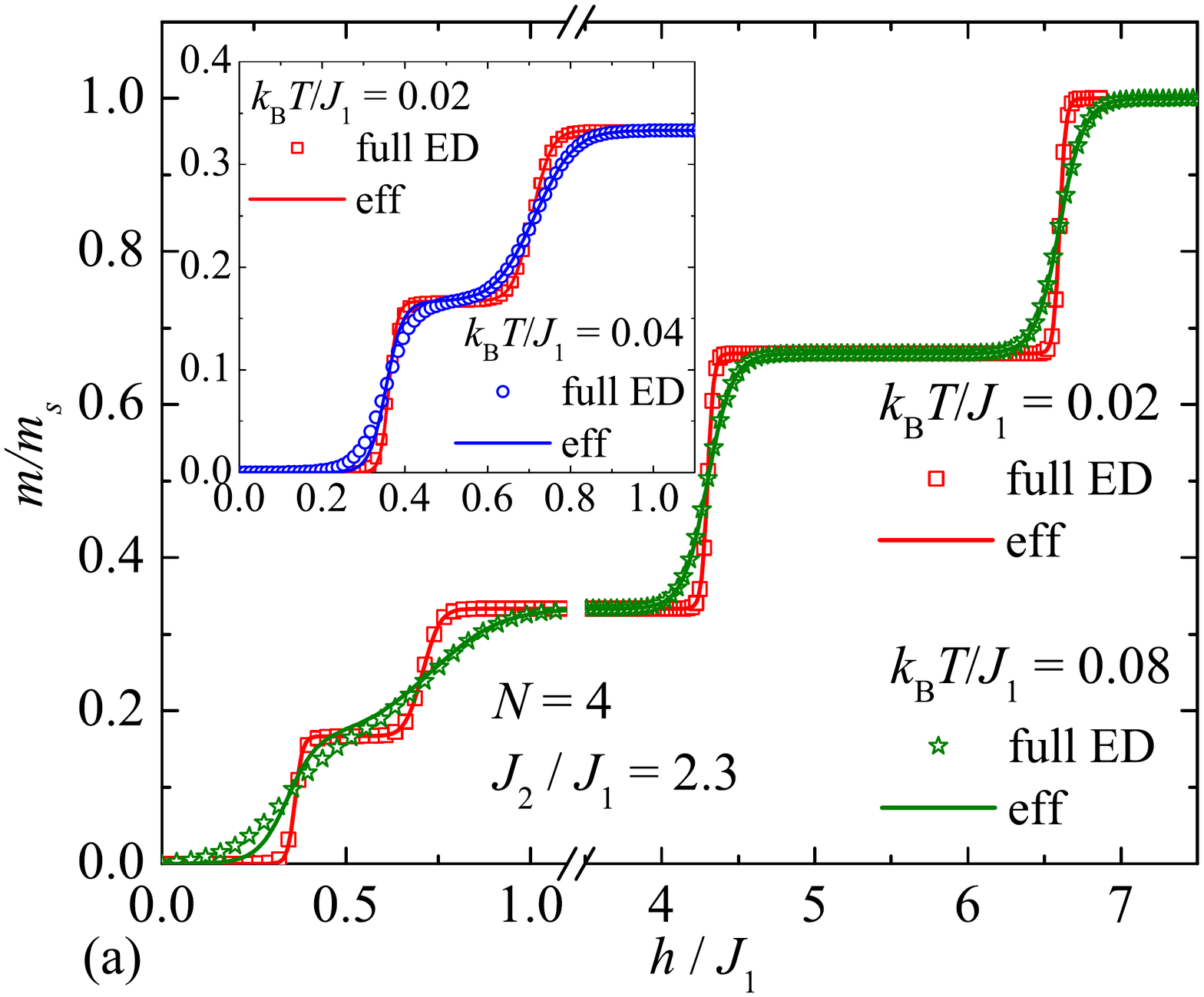}
\hspace{-1cm}
\includegraphics[width=0.5\textwidth]{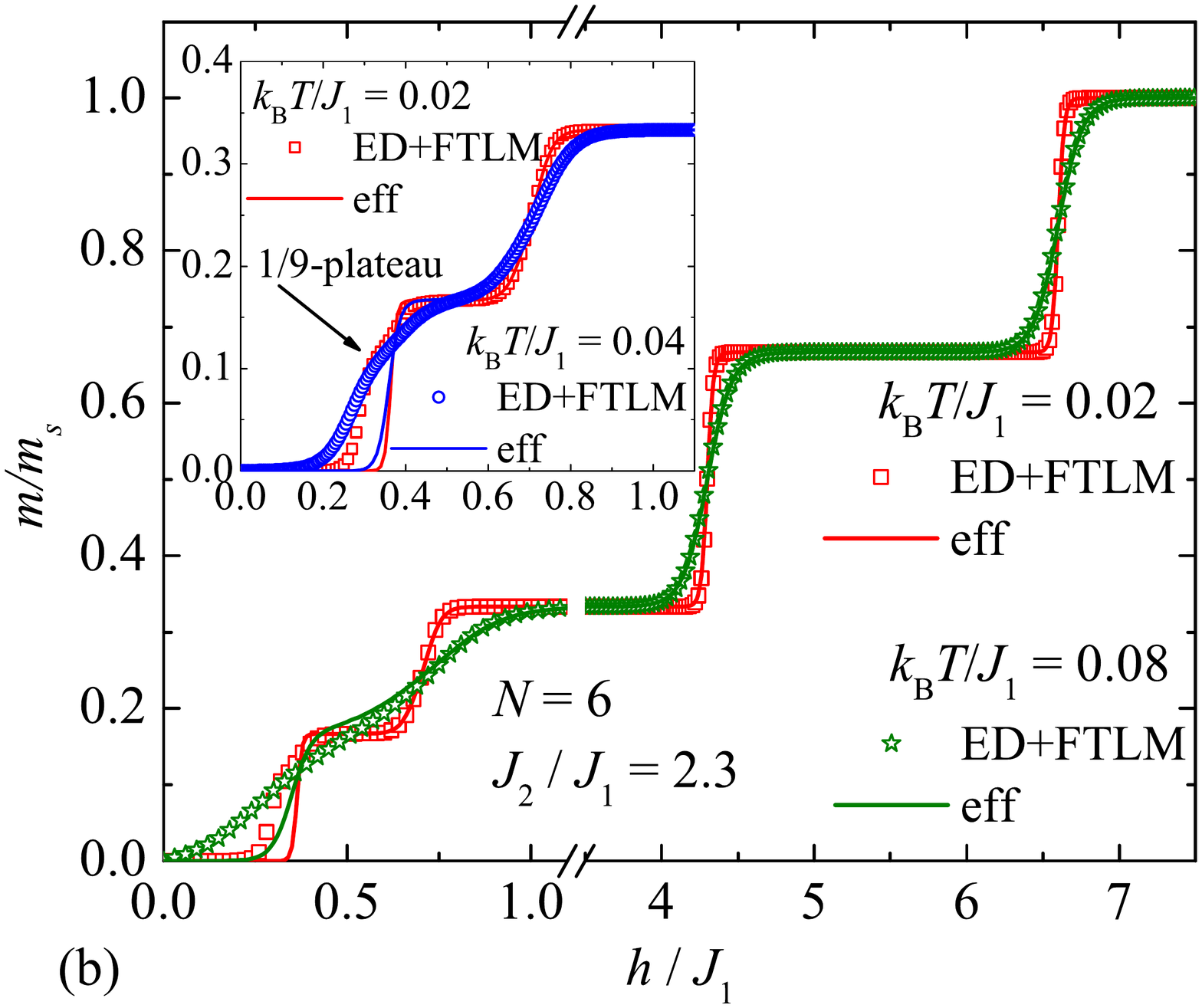}
\vspace{-1cm}
\includegraphics[width=0.5\textwidth]{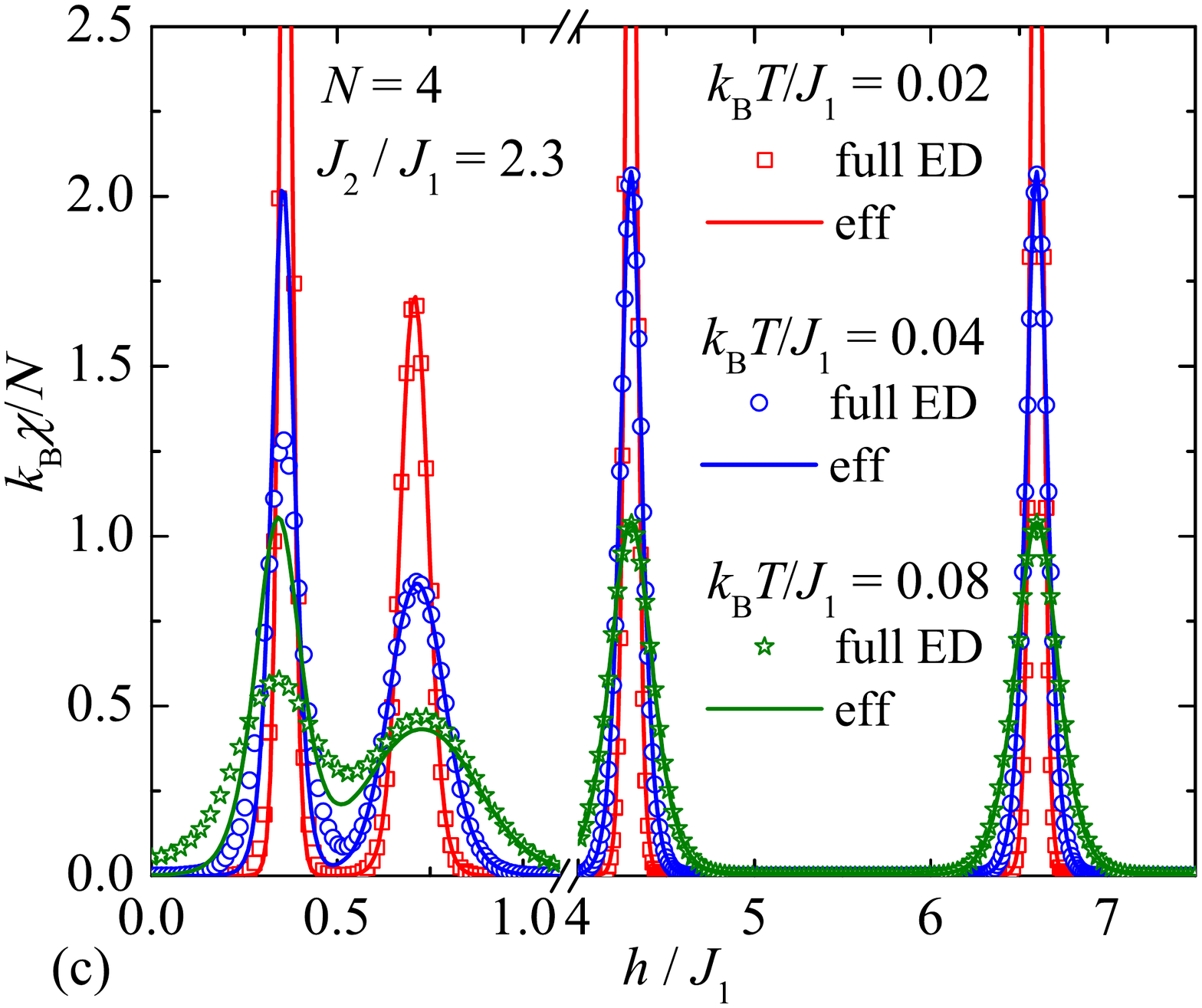}
\hspace{-1cm}
\includegraphics[width=0.5\textwidth]{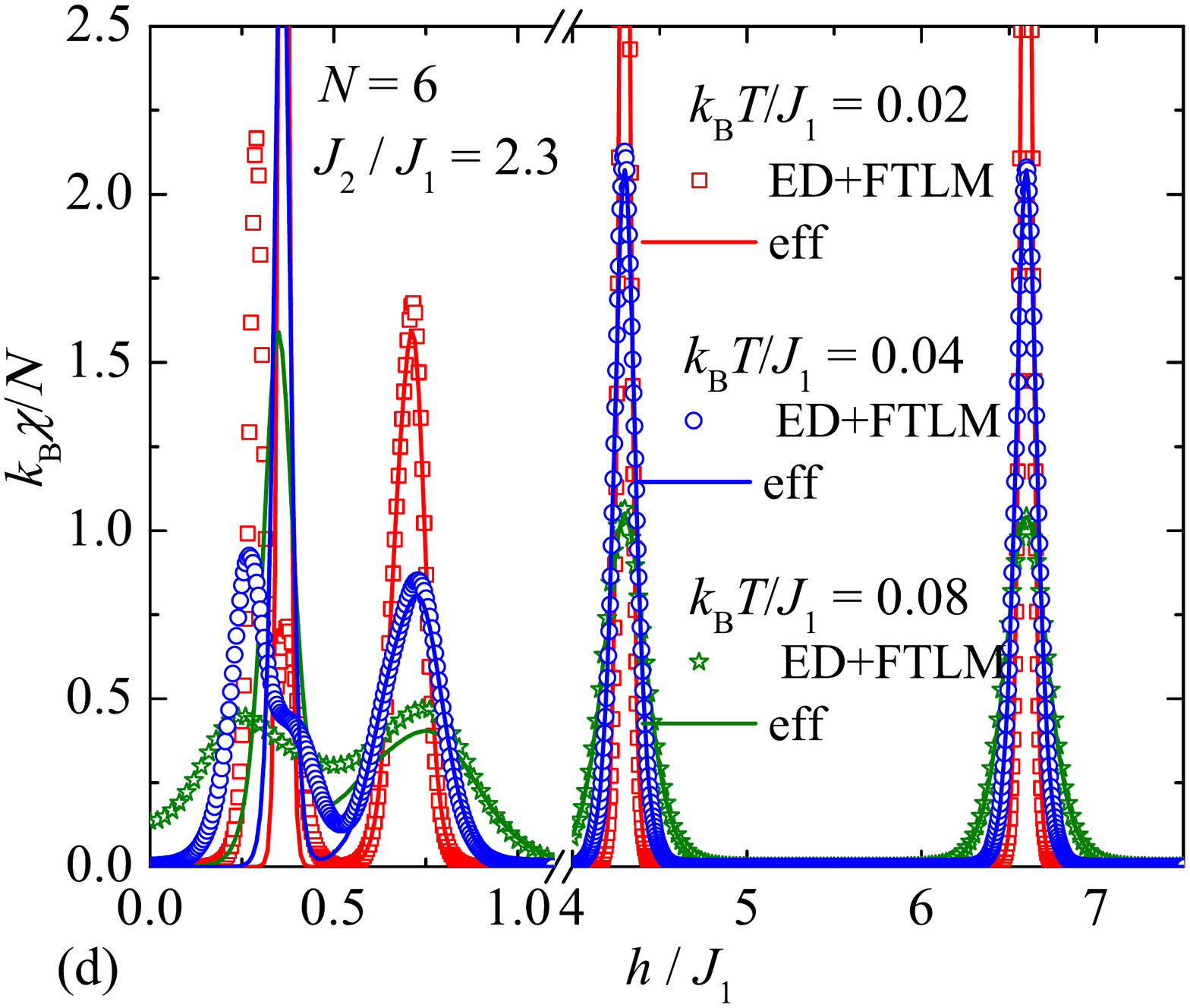}
\end{center}
\caption{A comparison of ED and FTLM data shown by symbols with the 
analytical results obtained from the effective lattice-gas model shown by 
solid lines for magnetic-field dependencies of (a)-(b) the magnetization and (c)-(d)
the susceptibility of the mixed-spin Heisenberg octahedral chain with $N=4$ (left panel) and $N=6$ (right panel) unit cells, the fixed value of the interaction ratio $J_2/J_1=2.3$ and a few selected values of temperature.
The magnetization is normalized with respect to its saturation value and susceptibility is normalized per unit cell.}
\label{fig3}       
\end{figure*}

The magnetic-field dependence of the entropy and the specific heat normalized per unit cell of the mixed-spin Heisenberg octahedral chain is shown in Fig. \ref{fig2} for $N=4$ and $N=6$ elementary unit cells, the fixed value of the interaction ratio $J_2/J_1=2.7$ and a few selected values of temperature. The entropy of the mixed-spin Heisenberg octahedral chain displays
a peak at each field-driven phase transition, while the specific heat exhibits
a double-peak behavior in the vicinity of all field-driven phase transitions. It is clear that
the height of all peaks in the entropy and the specific heat is invariant with respect to a small temperature change around the latter two critical fields, but it changes significantly around a first critical field.  It should be stressed that the entropy and
the specific heat are more sensitive for low-lying excitations above the 1/6-plateau neglected in the  effective monomer-dimer lattice-gas model, which can be seen by more substantial  discrepancies between numerical ED data and the analytical results obtained from the effective model. As a matter of fact, the more sizable differences are visible in
the low-field region even at lower temperature $k_{\rm B}T/J_1=0.04$ and become higher for higher temperature $k_{\rm B}T/J_1=0.08$ for both system sizes $N=4$ and $N=6$. Nevertheless, 
the effective monomer-dimer lattice-gas model qualitatively describes the peak behavior of
the entropy and the double-peak behavior of the specific heat even in the low-field region up to 
moderate temperatures $k_{\rm B}T/J_1\lesssim 0.08$. Moreover, the developed monomer-dimer lattice-gas model shows in
the high-field region a perfect agreement up to relatively high temperatures $k_{\rm B}T/J_1\approx 0.1$. This fact was comprehensively discussed in our previous work focusing 
on the description of the highly frustrated parameter region, where the effective lattice-gas model of two kinds of hard-core monomers satisfactorily captures
the 1/3- and 2/3-plateau as well as the low-temperature thermodynamics above them \cite{12karl19}.

\begin{figure*}
\begin{center}
\vspace{-1cm}
\includegraphics[width=0.5\textwidth]{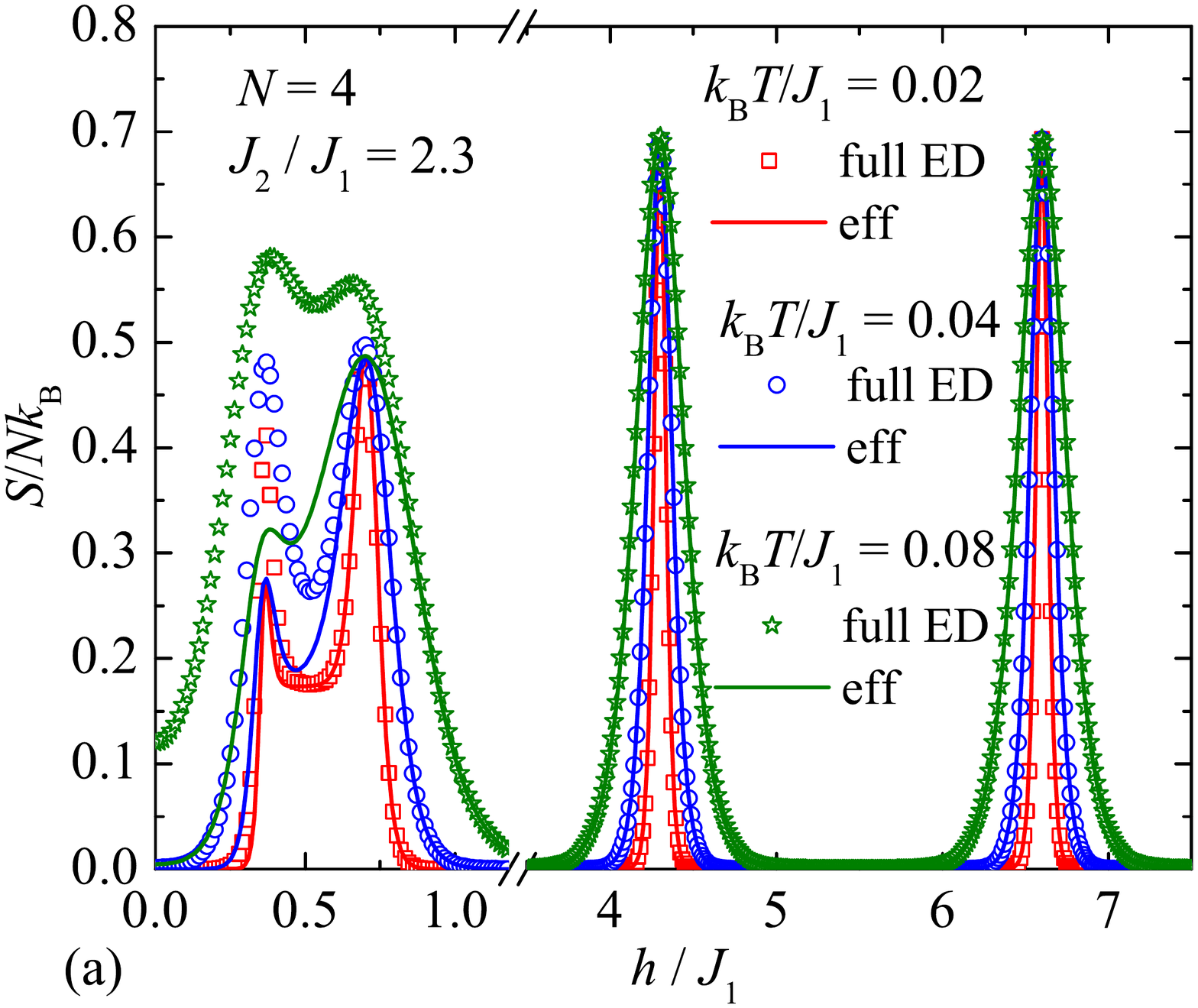}
\hspace{-1cm}
\includegraphics[width=0.5\textwidth]{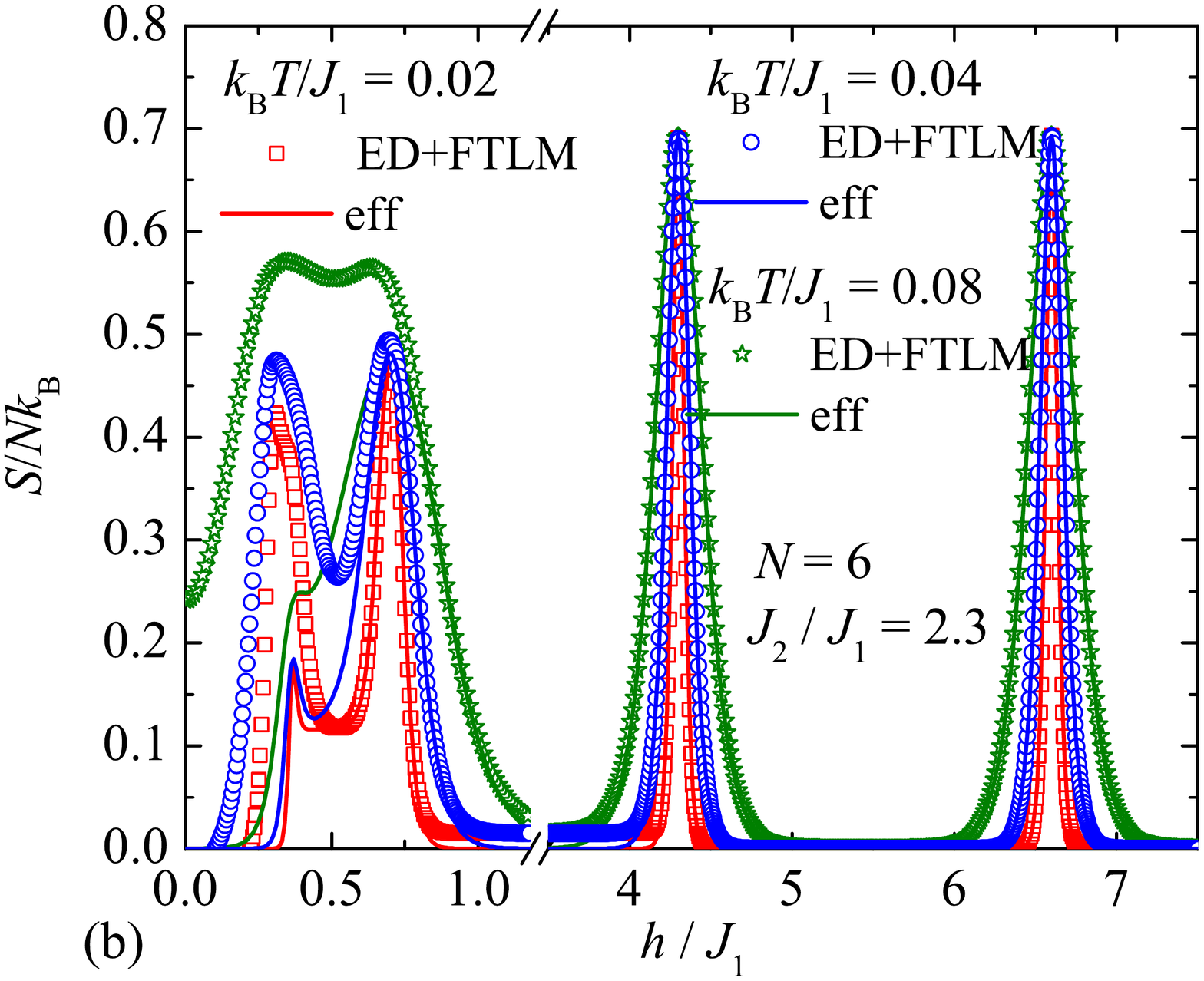}
\vspace{-1cm}
\includegraphics[width=0.5\textwidth]{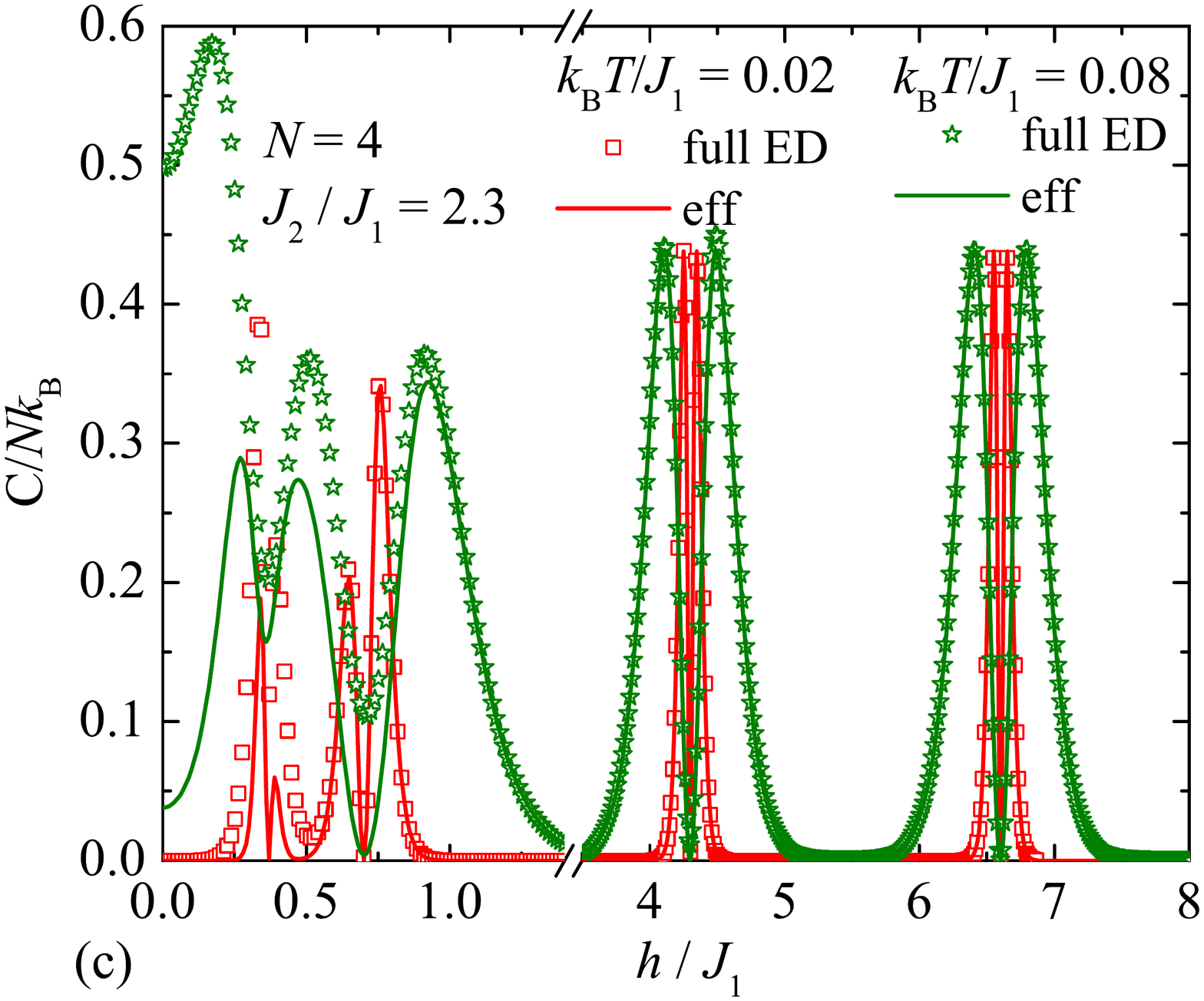}
\hspace{-1cm}
\includegraphics[width=0.5\textwidth]{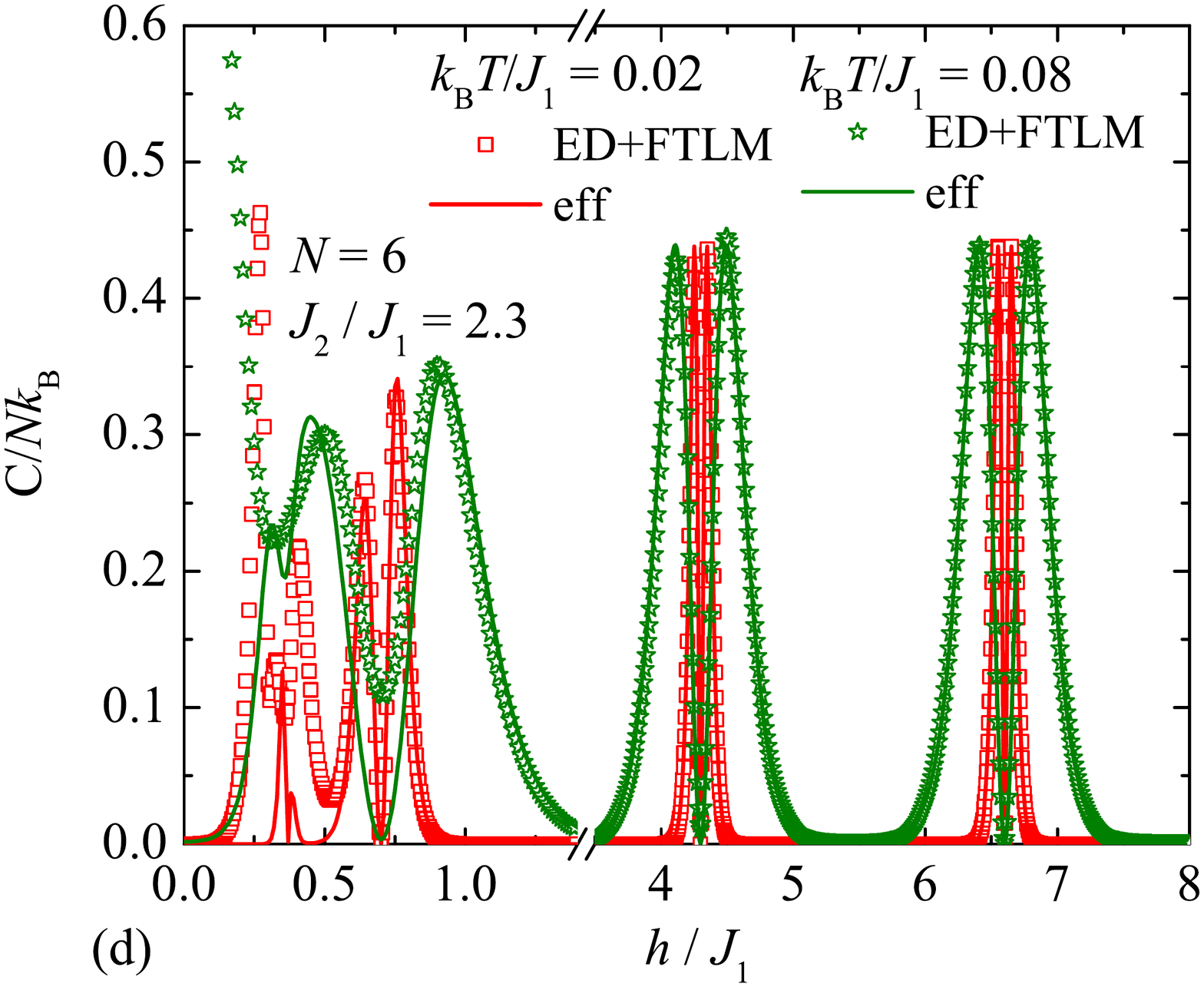}
\end{center}
\caption{A comparison of ED and FTLM data shown by symbols with the analytical results obtained from the effective monomer-dimer 
lattice-gas  model shown by solid lines for field dependencies of (a)-(b) the entropy and (c)-(d)
the specific heat of the mixed-spin Heisenberg octahedral chain with $N=4$ (left panel) and $N=6$ (right panel) unit cells,  the fixed value of the interaction ratio $J_2/J_1=2.3$ and a few selected values of temperature. Both quantities are normalized per unit cell.}
\label{fig4}       
\end{figure*}

In order to examine the less frustrated parameter region we have plotted in 
Figs. \ref{fig3} and \ref{fig4}  the same set of physical quantities of the mixed spin-(1,1/2) Heisenberg octahedral chain for $N=4$ (left panels) and $N=6$ (right panels) unit cells, the fixed value of the interaction ratio $J_2/J_1=2.3$ and a few different values of temperature.  In this parameter region, the mixed spin-(1,1/2) Heisenberg octahedral chain additionally displays the Haldane phase and  the magnetization curve shown in Figs. \ref{fig3}(a) and (b) indeed exhibits besides
the 2/3-, 1/3- and 1/6-plateaus a zero magnetization plateau, which is gradually blurred with increasing of the temperature. It can be seen from Fig. \ref{fig3}(a) that  the effective monomer-dimer lattice-gas model coincides well with the full ED data for four unit cells at  low enough temperatures $k_{\rm B}T/J_1\lesssim 0.04$, while there are more pronounced discrepancies between numerical and analytical results in
the low-field region at higher temperature $k_{\rm B}T/J_1= 0.08$. It should be noticed that
the magnetization curves shown in Fig. \ref{fig3}(b) for six unit cells show much greater discrepancy between the analytical results obtained from the  effective monomer-dimer lattice-gas model and numerical data obtained by combining the full ED with FTLM (ED+FTLM) due to presence of other cluster-based Haldane state manifested as the 1/9-plateau, which is neglected within the effective monomer-dimer lattice-gas model and cannot appear in the mixed-spin Heisenberg octahedral chain with the smaller number of unit cells $N=4$ due to insufficient system size required for this higher-period ground state.

It is quite obvious from the magnetic-field dependencies of the susceptibility shown in Fig.\ref{fig3}(c)  that analytical results obtained from monomer-dimer lattice-gas effective model for $N=4$ unit cells are in excellent agreement with full ED data above magnetic field $h/J_1>0.5$ up to relatively high temperatures $k_{\rm B}T/J_1\lesssim 0.08$, while they start to deviate more significantly at smaller magnetic fields $h/J_1\lesssim 0.5$. The effective monomer-dimer lattice-gas model 
overestimates the height of the susceptibility peak near the first field-induced transition between the Haldane and tetramer-hexamer phases, 
while it underestimates the local minimum between the first and second critical field $h/J_1\approx 0.5$ in comparison with precise numerical results. 
This disagreement is caused by the construction of the effective monomer-dimer lattice-gas model, which neglects low-lying excitations above the Haldane state associated mainly with
the emergence of other fragmentized cluster-based Haldane phases with higher periods, which are fully missing in the present effective monomer-dimer lattice-gas model that would need to be supplemented by hard-core trimeric and tetrameric particles. It actually turns out that
the presence of other cluster-based Haldane phase with period three being responsible for 1/9-plateau manifests itself in the magnetic-field dependence of the susceptibility shown in Fig. \ref{fig3}(d) for the mixed-spin octahedral chain with $N=6$ unit cells as additional peak with the maximum emergent approximately around magnetic field $h/J_1\approx 0.356$. This fact causes
the quantitative disagreement of
the height of the low-field peaks as well as the qualitative discrepancies between results obtained from the effective monomer-dimer lattice-gas model and numerical calculations based 
on the combination of full ED and FTLM. The disagreement concerns thus with the position of the first peak and
the absence of
a peak corresponding to 1/9-plateau in the analytical results obtained from the effective monomer-dimer lattice-gas model.

The  magnetic-field dependence of the entropy and the specific heat of the mixed spin-(1,1/2) Heisenberg octahedral chain with $N=4$ and $N=6$ unit cells are depicted in Fig. \ref{fig4} for the interaction ratio $J_2/J_1=2.3$. It is evident from Fig. \ref{fig4}(a)-(b) that the entropy exhibits sharp peaks in the proximity of all critical fields at very low temperature $k_{\rm B}T/J_1=0.02$, which become smeared out upon increasing of temperature. The results obtained from the effective monomer-dimer lattice-gas model and the full ED data for $N=4$ or by combining the full ED with FTLM for $N=6$, are in reasonable agreement for magnetic fields higher than $h/J_1\gtrsim 0.75$ up to moderate temperatures $k_{\rm B}T/J_1\approx 0.08$, while the observed deviations in
the low-field region are mainly caused by the neglecting low-lying excited states above the Haldane phase in the effective monomer-dimer lattice-gas model. 

On the other hand, the specific heat as a function of the magnetic field of the mixed spin-(1,1/2) Heisenberg octahedral chain for $N=4$ and $N=6$ unit cells plotted in Fig. \ref{fig4}(c)-(d) 
exhibits a double-peak behavior near each field-driven phase transition for the interaction ratio $J_2/J_1=2.3$. It should be stressed that the analytical results obtained from the effective monomer-dimer lattice-gas model and the numerical results obtained for the specific heat from ED and FTLM are in a plausible accordance in
the high-field region $h/J_1\gtrsim 0.5$ up to relatively high temperatures $k_{\rm B}T/J_1\approx 0.08$, while
the disagreement at lower magnetic fields $h/J_1<0.5$ relates to neglecting low-lying excited states above the Haldane phase as well as neglecting the higher-period cluster-based Haldane ground state corresponding to the 1/9-plateau emergent for larger system size $N=6$ [see Fig. \ref{fig4}(d)].

\begin{table}
\caption{Zero-field energy and degeneracy of the lowest-energy eigenstates  with the given $z$-component of the total spin $S_T^z$ ranging between fully polarized state $S_T^z=18$ and the eigenstate corresponding to the 2/3-plateau of the mixed spin-(1,1/2) Heisenberg octahedral chain with $N=6$ unit cells (30 spins) by assuming two different values of the interaction ratio $J_2/J_1=2.3$ and $J_2/J_1=2.7$. The last column determines the total number of magnons.}
\begin{center}
\begin{tabular}{ |p{0.5cm}||p{1.5cm}|p{1.5cm}|p{0.5cm}|p{0.3cm}| }
 \hline
$S_T^z$& $J_2/J_1$=2.3&$J_2/J_1$=2.7& deg& $n$\\
 \hline
 18 & +37.8   & +40.2     & 1&0\\
 17 & +31.2   & +32.8     & 6&1  \\
 16 & +24.6   & +25.4     & 15&2\\
 15 & +18.0   & +18.0     & 20&3\\
 14 & +11.4   & +10.6     & 15&4\\
 13 & +4.8    & +3.2      & 6&5  \\
 12 & $-1.8$  & $-4.2$    & 1&6\\
\hline
\end{tabular}
\end{center} 
\label{tabulka}
\end{table}
\begin{figure*}
\begin{center}
\vspace{-1cm}
\includegraphics[width=0.5\textwidth]{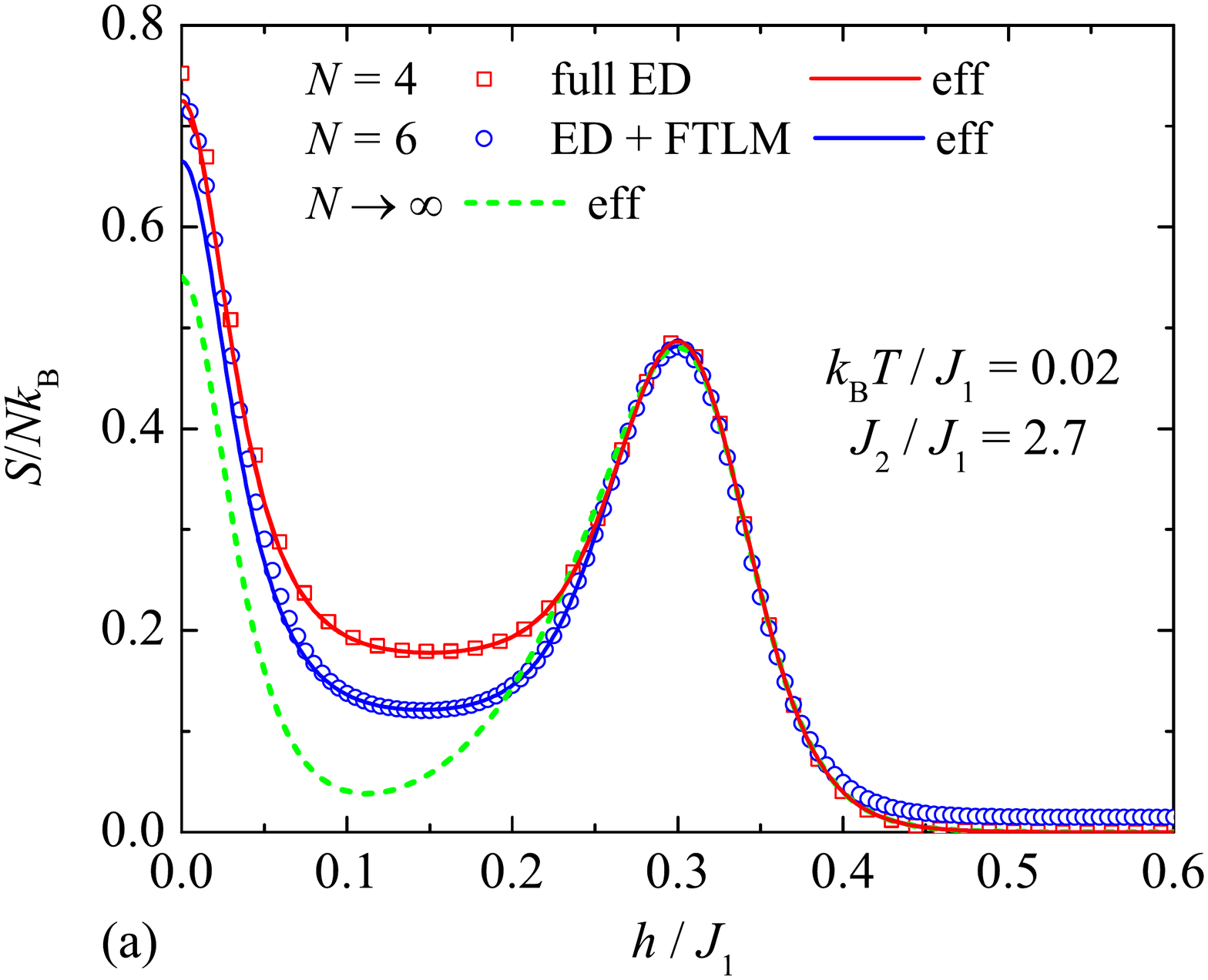}
\hspace{-1cm}
\includegraphics[width=0.5\textwidth]{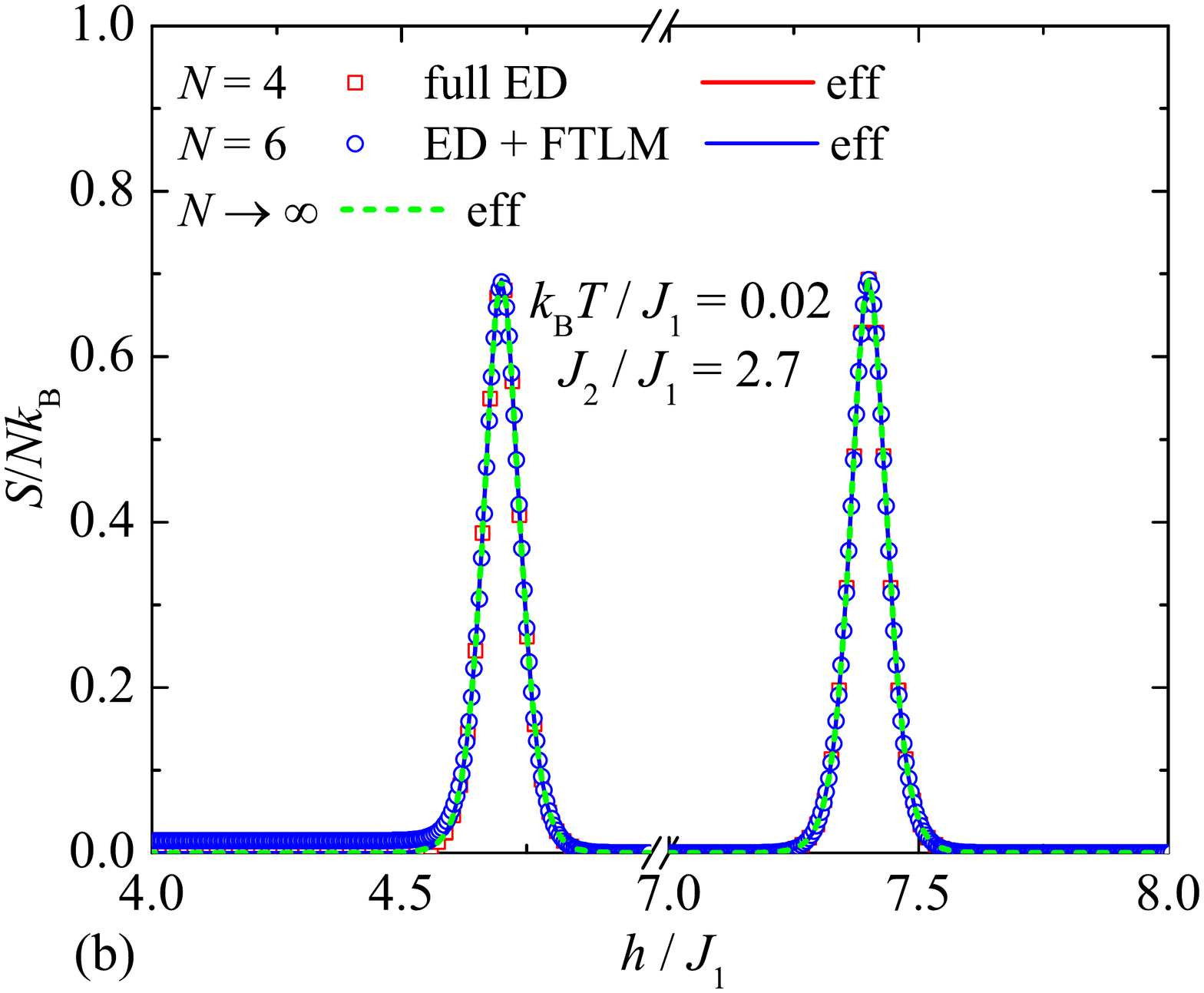}
\vspace{-1cm}
\includegraphics[width=0.5\textwidth]{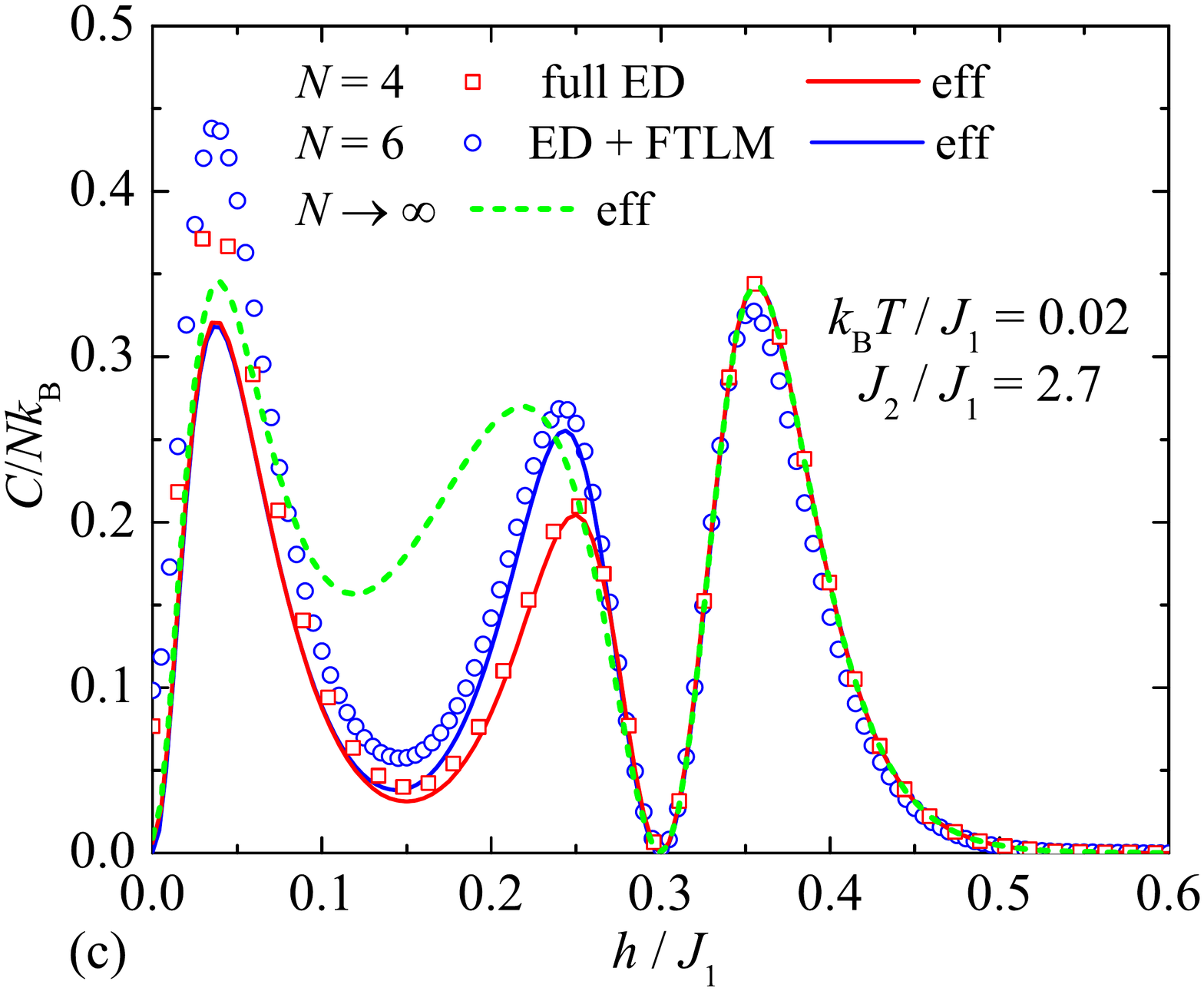}
\hspace{-1cm}
\includegraphics[width=0.5\textwidth]{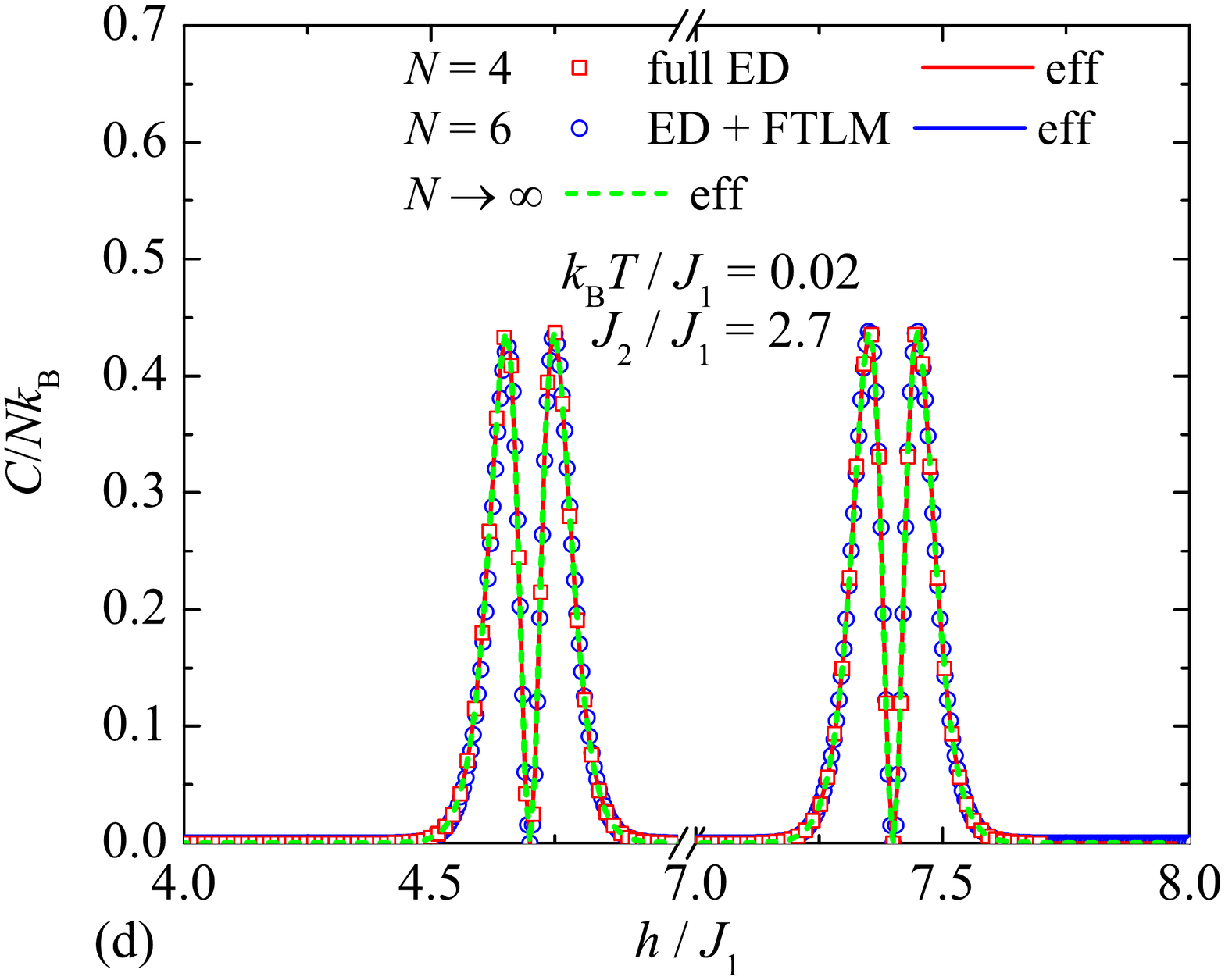}
\end{center}
\caption{Magnetic-field dependencies of the entropy  (a)-(b) and the specific heat (c)-(d) of the mixed spin-(1,1/2) Heisenberg octahedral chain at low temperature $k_{\rm B}T/J_1=0.02$ for three different system sizes with $N=4$, $N=6$ and $N\to\infty$ unit cells. 
There are obvious finite-size effects in the entropy and specific heat at low fields (left panel), but the finite-size effects are totally absent in
the high-field region (right panel).}
\label{fig5}       
\end{figure*}

Energies and degeneracies of the lowest-energy eigenstates of the mixed spin-(1,1/2) Heisenberg octahedral chain with $N=6$ are presented in Tab. \ref{tabulka} for a few selected values of the total spin $S_T^z=12,13,...,18$. The fully polarized ferromagnetic state with $S_T^z=18$ is nondegenerate, while the degeneracy of the state with $n$-spin deviations from the fully polarized ferromagnetic state equals to the combinatorial number $6\choose{n}$. The respective degeneracy $6\choose{n}$ relates to the number of all available combinations with $n$ localized magnons placed on 6 square plaquettes of the octahedral chain. For instance, the lowest-energy eigenstate from sector with $S_T^z=17$ corresponds to the state, in which all spins are polarized except one square plaquette involving one localized magnon ($n=1$), etc. The lowest-energy eigenstate from the last sector with $S_T^z=12$ corresponds to the bound magnon-crystal ground state (\ref{BM}), which has one localized magnon on each square plaquette and is responsible for magnon crystallization within 2/3-plateau.

To bring deeper insight into the finite-size effects of the mixed spin-(1,1/2) Heisenberg octahedral chain, we have displayed in Fig. \ref{fig5}
the field dependence of the entropy and the specific heat at relatively small temperature $k_{\rm B}T/J_1=0.02$ with three different lattice sizes $N=4$, $N=6$ and $N\to\infty$ and the interaction ratio $J_2/J_1=2.7$. 
A comparison between the obtained analytical and numerical results, as well as behavior of these quantities was comprehensively discussed 
above, let us therefore focus our attention to finite-size effects only. It can be seen from Fig. \ref{fig5} that sizable finite-size effects are present in the entropy and specific heat only at low fields (left panel in Fig. \ref{fig5}), while they are almost 
totally absent in
the high-field region (right panel in Fig. \ref{fig5}). More specifically, the entropy differs for different system sizes from zero field nearly up to the magnetic-field value $h/J_1\approx 0.25$. 
The higher the system size is, the lower is the zero-field entropy. Similarly, finite-size effects of the specific heat persists up to nearly the same value of the magnetic field $h/J_1\approx 0.25$ (see Fig. \ref{fig5}(c)). 

\begin{figure*}
\begin{center}
\vspace{-1cm}
\includegraphics[width=0.5\textwidth]{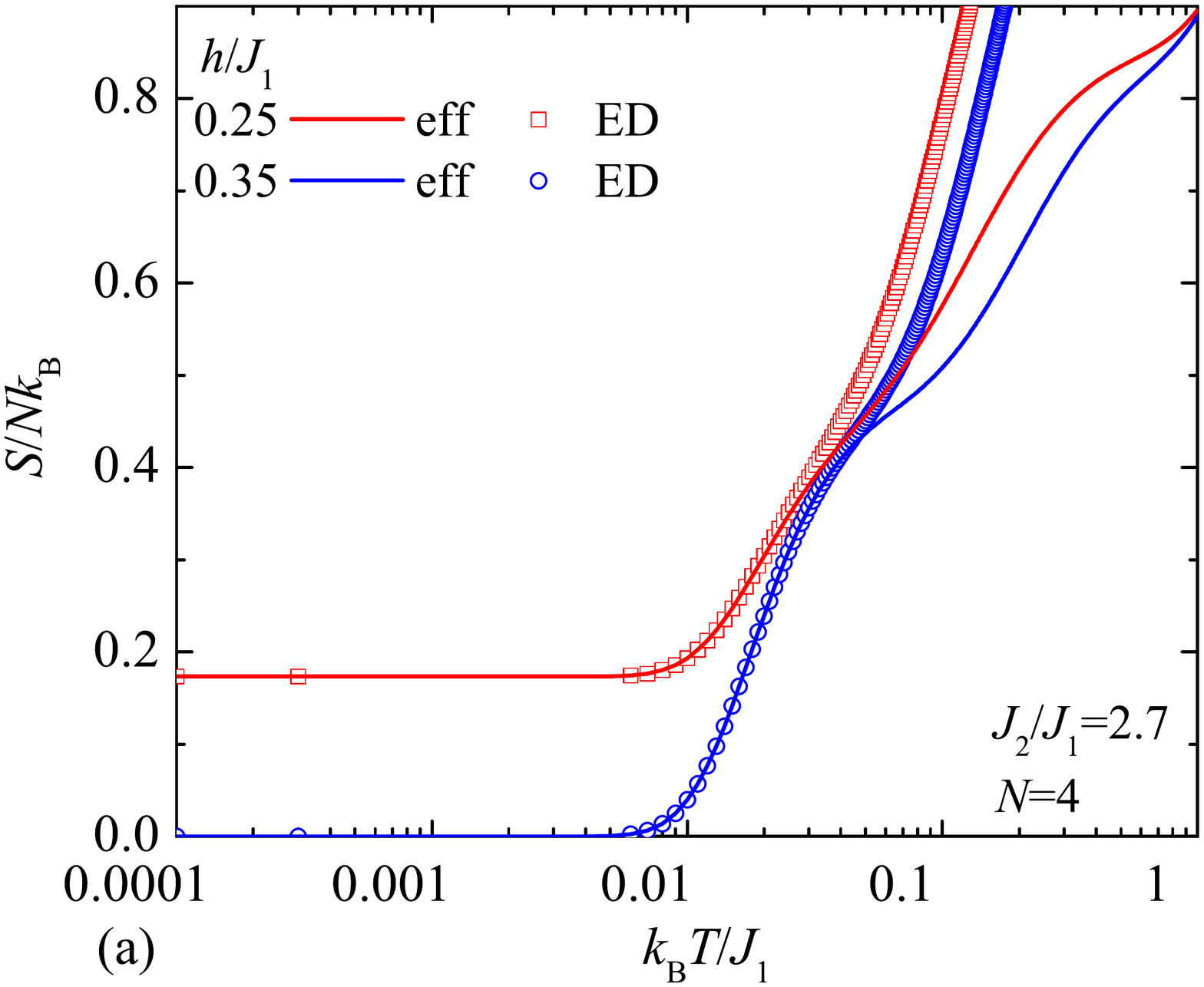}
\hspace{-1cm}
\includegraphics[width=0.5\textwidth]{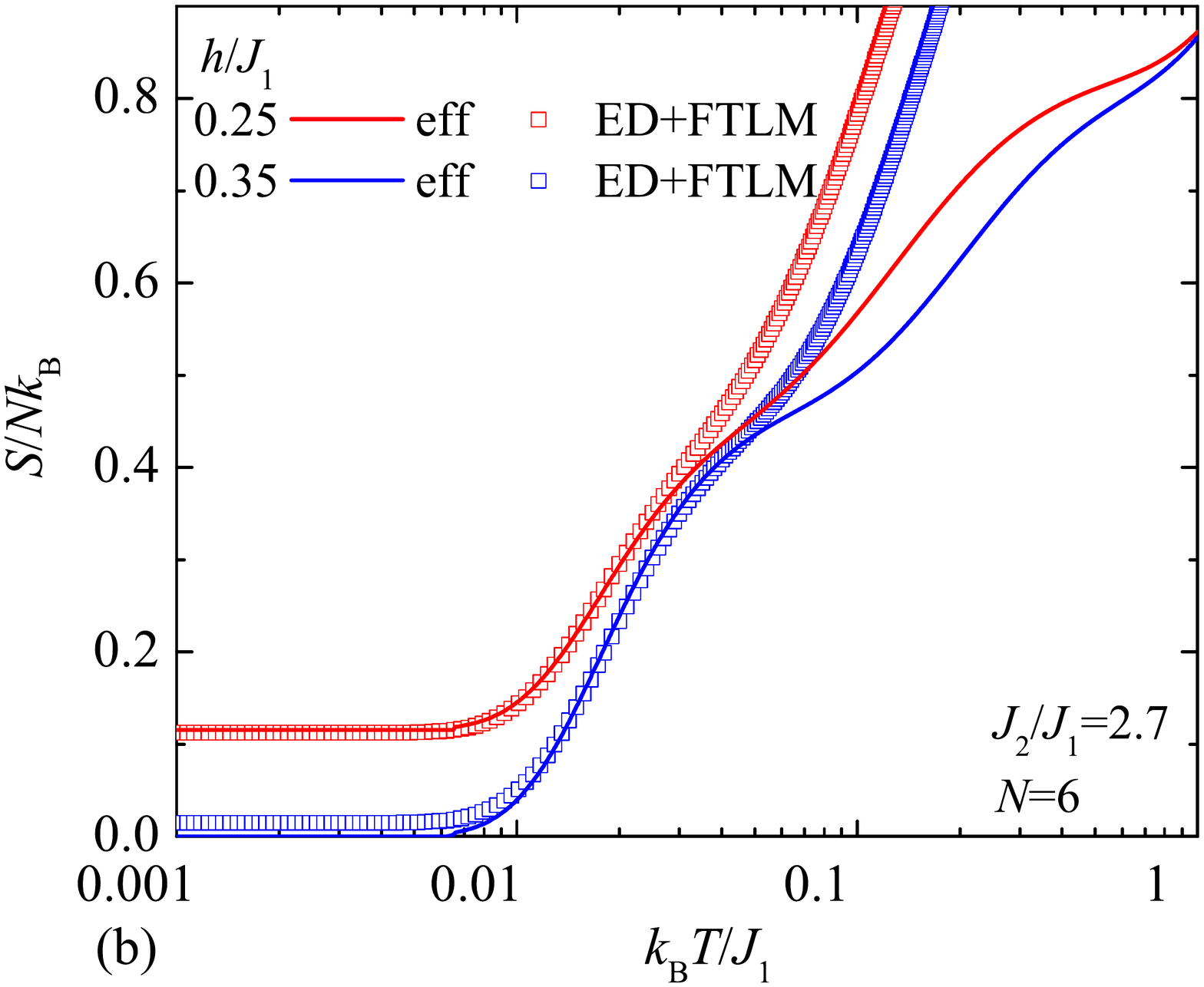}
\includegraphics[width=0.5\textwidth]{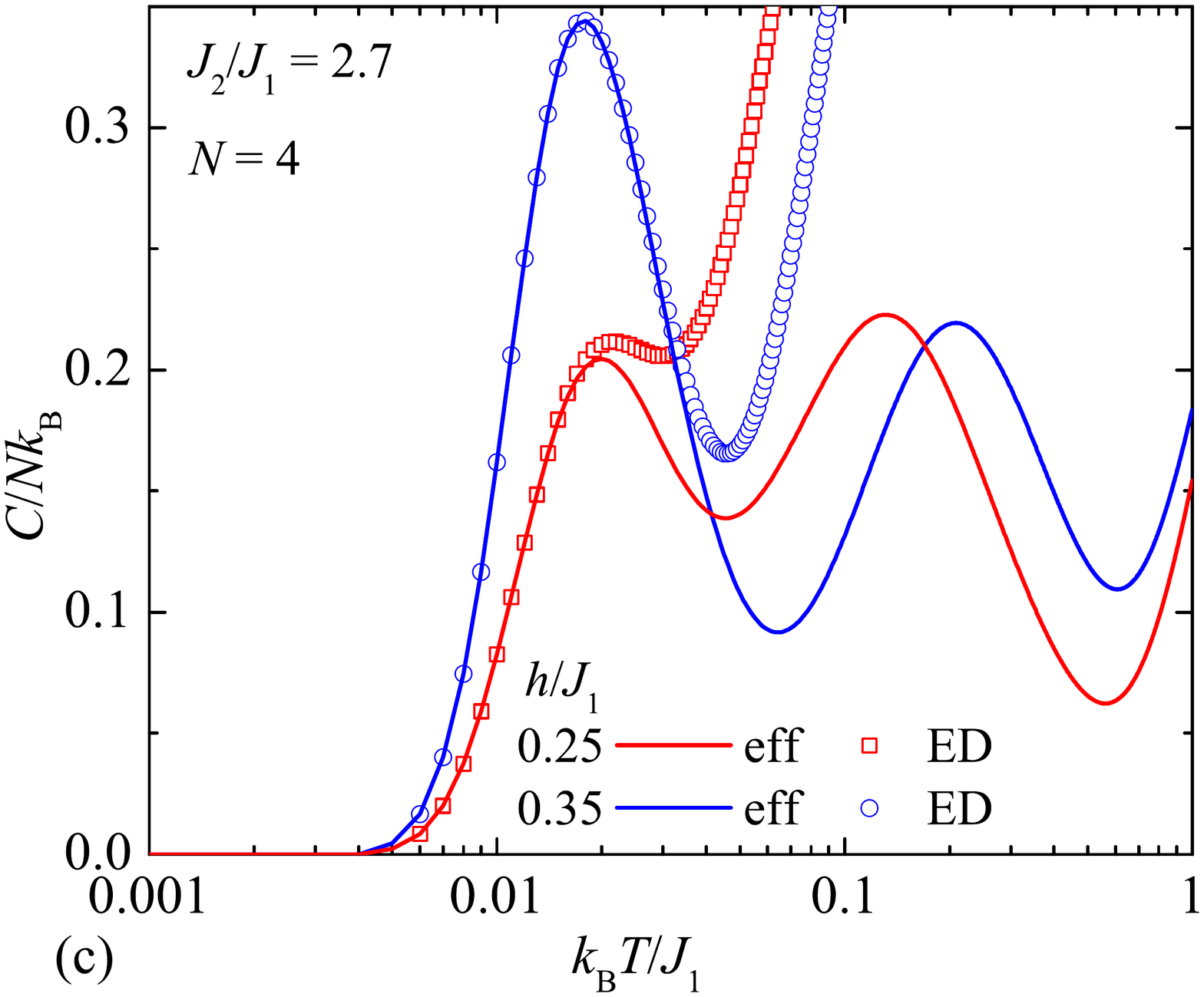}
\hspace{-1cm}
\includegraphics[width=0.5\textwidth]{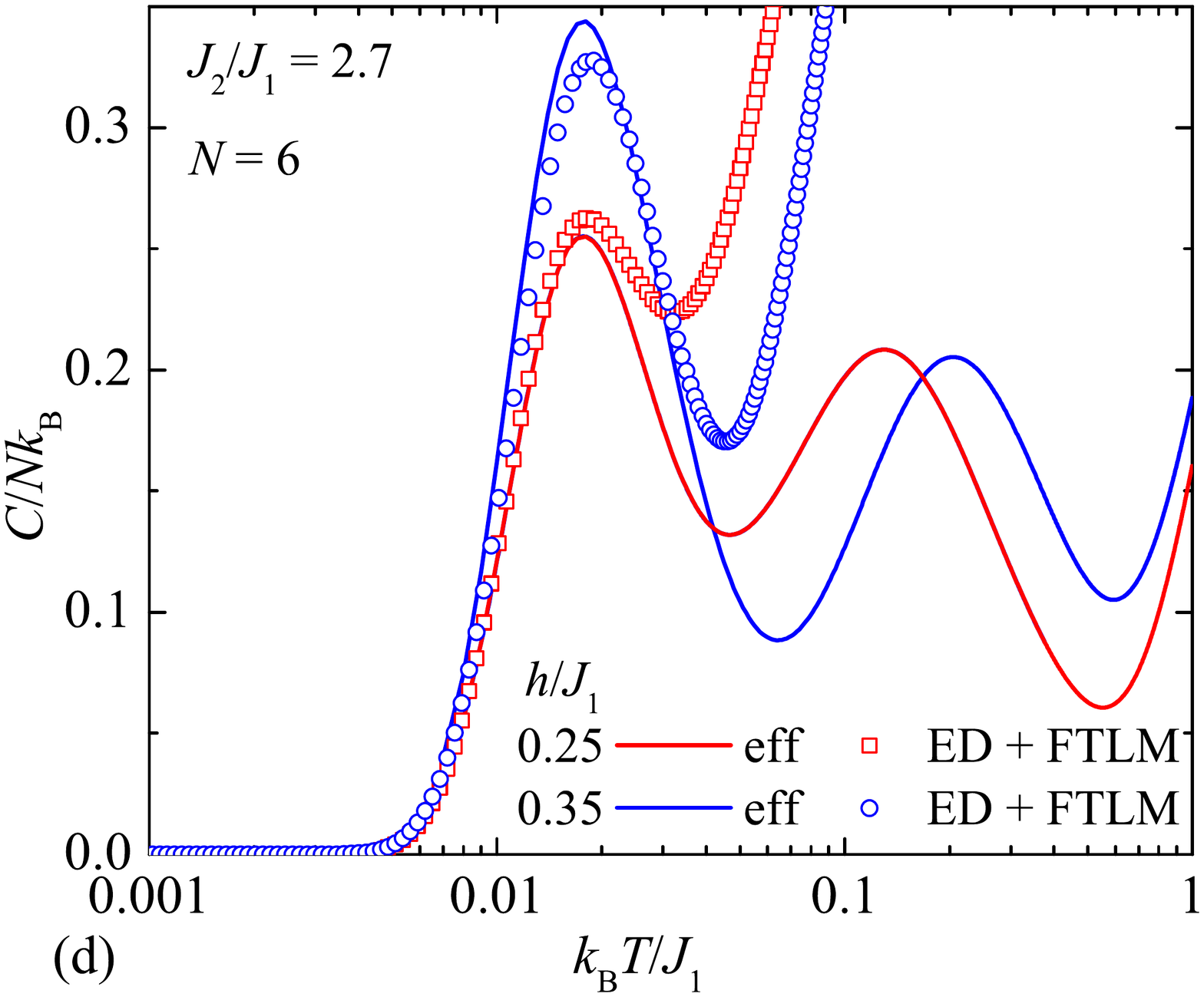}
\end{center}
\vspace{-0.8cm}
\caption{Temperature dependencies of the entropy (a)-(b) and specific heat (c)-(d) of the mixed spin-1 and 1/2 Heisenberg octahedral chain with $N=4$ (left panel) and $N=6$ (right panel) unit cells and the interaction ratio $J_2/J_1=2.3$ just below ($h/J_1=0.25$) and above ($h/J_1=0.35$) the critical field $h_c/J_1=0.3$.  At the critical field  monomer-tetramer phase and tetramer-hexamer phase of mixed-spin 1/2 and spin 1 Heisenberg octahedral chain coexist together.}
\label{fig6}       
\end{figure*}

Finally, let us discuss the temperature dependencies of the entropy and the specific heat of the mixed spin-(1,1/2) Heisenberg octahedral chain with $N=4$ and $N=6$ unit cells at the interaction ratio $J_2/J_1=2.7$ just below and just above the first critical field $h_c/J_1=0.3$, as displayed in Fig. \ref{fig6}. If the magnetic field is set to the value $h/J_1=0.25$, the ground state of the mixed spin-(1,1/2) Heisenberg octahedral chain is two-fold degenerate tetramer-hexamer phase. Indeed, the entropy asymptotically matches in the limit of zero temperature the value $S/Nk_{\rm B}=\ln 2/4 \doteq 0.172$ for $N=4$ and $S/Nk_{\rm B}=\ln 2/6 \doteq 0.116$ for $N=6$ unit cells (see Fig. \ref{fig6}(a) and (b), respectively). On the other hand, the ground state of the mixed spin-(1,1/2) Heisenberg octahedral chain is for the other higher field value $h/J_1=0.35$ the nondegenerate monomer-tetramer phase and thus, the zero-temperature entropy tends trivially to zero for both system sizes ($N=4$ and $N=6$). In any case, the entropy persist almost constant also at finite temperatures 
up to the value $k_{\rm B}T/J_1 \approx 0.01$ when it starts to rise with the further increase of temperature.  From the comparison of the analytical results derived from the effective monomer-dimer lattice-gas model with ED data for $N=4$ unit cells one may conclude that the effective monomer-dimer lattice-gas 
model correctly predicts the behavior of the entropy whenever the temperature is lower then $k_{\rm B}T/J_1\lesssim 0.05$, while analytical and numerical data in Fig. \ref{fig6}(a) starts to deviate above this temperature. The effective monomer-dimer lattice-gas model almost copies data obtained from the full ED and FTLM method also for the entropy of the mixed spin-(1,1/2) Heisenberg octahedral chain with $N=6$ unit cells up to $k_{\rm B}T/J_1\lesssim 0.03$ (see Fig. \ref{fig6}(b)). 
Note that the tiny deviation of  the numerical FTLM data for the entropy from zero 
as $T \to 0$ can be attributed to the approximate nature of the FTLM.

Temperature dependencies of the specific heat  of the mixed spin-(1,1/2) Heisenberg octahedral chain with $N=4$ and $N=6$ unit cells  are plotted in Fig. \ref{fig6}(c) and (d) for the interaction ratio $J_2/J_1=2.7$ and two magnetic fields slightly 
above and slightly below the field-induced phase
transition between the tetramer-hexamer and monomer-tetramer phase. The specific heat is zero up to $k_{\rm B}T/J_1\approx 0.005$ and then it displays
a striking temperature dependence with 
a pronounced low-temperature maximum.  
From the comparison of results obtained from the effective monomer-dimer lattice-gas model (solid lines) and numerical data acquired by ED and FTLM
(symbols) can be concluded that the effective monomer-dimer lattice-gas model qualitatively as well as quantitatively describes
the low-temperature peak of the specific heat. The validity of the effective description seems to be much better when considering the magnetic fields exceeding the first critical field, because the analytical and numerical data coincide up to higher temperatures.

\section{Concluding remarks}
\label{conclusion}
In the present work we have investigated in detail the mixed spin-(1,1/2) Heisenberg octahedral chain in presence of the external magnetic field using the extended version of the localized-magnon approach, which establishes a mapping relationship with the effective lattice-gas model of hard-core monomers and dimers. By the use of monomeric and dimeric particles we have afforded 
a classical description 
of the fully quantum mixed spin-(1,1/2) Heisenberg octahedral chain. We have compared our analytical results obtained from the effective monomer-dimer lattice-gas model with the numerical calculations obtained from the full ED and FTLM of the mixed spin-(1,1/2) Heisenberg octahedral chain for two system sizes with $N=4$ and $N=6$ unit cells and we have shown that the effective monomer-dimer 
lattice-gas model satisfactorily
describes the 2/3-, 1/3-, as well as 1/6-plateau and the zero magnetization plateau in
the moderately frustrated parameter regime, which involves the cluster-based Haldane phase
(with character of
a tetramer-hexamer ground state) and the uniform Haldane phase. Moreover, the effective monomer-dimer lattice-gas model 
qualitatively describes thermodynamic quantities of the mixed spin-(1,1/2) Heisenberg octahedral chain such as susceptibility, entropy and specific heat at very low temperatures and explains 
the degeneracy of the lowest-energy eigenstates with given $S_T^z$ ranging from
the fully polarized state up to the eigenstate corresponding to the 2/3-plateau.

Besides this, we have proven that sizable finite-size effects appear in
the low-field region, while there are no finite-size effects in the high-field region. The presented effective description is thus especially valuable, because it provides reasonable results
in the thermodynamic limit
not accessible to unbiased numerical methods. The theory based on the classical description developed from the monomeric and dimeric hard-core particles lacks low-lying excitations above the 
Haldane phase, which become relevant even at very low temperatures. We consider this issue as a future challenging task aimed at better quantitative description of thermodynamics of fully frustrated quantum spin systems including the uniform and higher-period cluster-based Haldane phases. 

\begin{acknowledgments}
This work was financially supported by the grant of The Ministry of Education, Science, Research and Sport of the Slovak Republic under the contract No. VEGA 1/0531/19 and by the grants of the Slovak Research and Development Agency under the contract No. APVV-20-0150. K.K. acknowledges kind hospitality during summer 2021 at Max-Planck-Institut
f\"ur Physik Komplexer Systeme in Dresden, where the major part of this work was completed.
J.R. thanks the Deutsche
Forschungsgemeinschaft for financial support (DFG  RI 615/25-1).
\end{acknowledgments}

\end{document}